\documentclass{cpbtex}
\usepackage{graphicx}
\usepackage{amssymb}
\usepackage{amsmath}
\usepackage{graphicx}
\usepackage{mathrsfs}
\usepackage{mdwlist}
\usepackage{subfigure}
\usepackage{booktabs}
\usepackage{amsmath}
\usepackage{extarrows}
\usepackage{dsfont}
\usepackage{amstext}
\usepackage{amsbsy}
\usepackage{bbm}
\usepackage{bm}
\usepackage{amssymb,mathrsfs,amsmath}
\usepackage{threeparttable}
\usepackage{booktabs}
\usepackage{bbding,pifont}
\usepackage{amsthm}
\usepackage{graphicx}
\usepackage{textcomp}
\usepackage{pifont}
\usepackage{url}
\usepackage{caption}

\begin{document}

\title{Effective dynamics and quantum state engineering by periodic kicks}

\author{Zhi-Cheng Shi$^{1,2}$, \ Zhen Chen$^{1,2}$, \ Jian-Hui Wang$^{1,2}$, \\ Yan Xia$^{1,2}$\thanks{Corresponding author. E-mail: xia-208@163.com}, \ and \ X. X. Yi$^{3}$\thanks{Corresponding author. E-mail: yixx@nenu.edu.cn}\\
$^{1}${Fujian Key Laboratory of Quantum Information and Quantum Optics}\\{ (Fuzhou University), Fuzhou 350108, China}\\  
$^{2}${Department of Physics, Fuzhou University, Fuzhou 350108, China}\\ 
$^{3}${Center for Quantum Sciences and School of Physics, Northeast Normal University,}\\{ Changchun 130024, China}}   


\maketitle

\begin{abstract}
In this work, we study the kick dynamics of periodically driven quantum systems, and provide a time-independent effective Hamiltonian with the analytical form to reasonably describe the effective dynamics in a long timescale. It is shown that the effective coupling strength can be much larger than the coupling strength of the original system in some parameter regions, which stems from the zero time duration of kicks. Furthermore, different regimes can be transformed from and to each other in the same three-level system by only modulating the period of periodic kicks. In particular, the population of excited states can be selectively suppressed in periodic kicks, benefiting from the large detuning regime of the original system. Finally, some applications and physical implementation of periodic kicks are demonstrated in quantum systems. Those unique features would make periodic kicks becoming a powerful tool for quantum state engineering.
\end{abstract}

\textbf{Keywords:} periodic driving, $\delta$ function, quantum state engineering

\textbf{PACS:} 42.50.-p, 42.50.Dv, 02.30.Yy

\section{Introduction}

Periodically driven systems have attracted a lot of attention in recent years.
Unlike the static system, periodically driven systems do not have stationary states, but possess the well-defined quasi-stationary states explained by the Floquet theory.\ucite{PhysRev.138.B979,PhysRevA.7.2203} By regulating its quasi-energy spectrum, many non-trivial phenomena are predicted to appear in those systems. The celebrated one is the coherent destruction of tunneling phenomenon,\ucite{PhysRevLett.67.516,Grifoni1998} in which the populations of quantum states are suppressed via properly modulating the parameters of the driving field.
This novel phenomenon can be understood by the destructive quantum interference.
Furthermore, some typical periodically driven systems, such as ultracold atomic
gases in a harmonic potential and the driven spin chains with Ising symmetry, are lately exploited to search for quantum time crystals,\ucite{PhysRevA.91.033617,PhysRevLett.116.250401,PhysRevLett.117.090402} which have spontaneous breaking of time-translation symmetry.\ucite{PhysRevLett.109.160401,Sacha_2017}

Until now, the utility of periodic driving fields in quantum systems is still a popular and intuitive way of gaining insight into some novel physical effects,
e.g., dynamical localization,\ucite{PhysRevA.76.043625,PhysRevB.90.174407,PhysRevA.91.063607, Bukov2015,PhysRevB.97.020303,PhysRevA.98.013635,PhysRevB.97.100301} quantum phase transition,\ucite{PhysRevLett.95.260404,PhysRevLett.99.196403,PhysRevB.90.214502,
RevModPhys.89.011004,PhysRevLett.127.250402,PhysRevB.105.054205} and multiphoton resonances suppression.\ucite{PhysRevLett.119.053203}
Moreover, periodic driving fields have also been regarded as a convenient means for quantum coherent control,\ucite{PhysRevLett.102.100403,PhysRevLett.105.086804,PhysRevLett.109.210501,PhysRevLett.113.193601,PhysRevA.91.052122,PhysRevLett.122.120403}
including special quantum states manipulation\ucite{PhysRevLett.99.110501,PhysRevB.78.064503,Li_2009,Shi2018,PhysRevA.104.053101} and
typical quantum gates implementation.\ucite{PhysRevB.74.140504,PhysRevA.94.012321,PhysRevA.100.022337,PhysRevA.101.042314,PhysRevA.103.012601}

There are several forms of periodic driving fields, such as sinusoidal forms, square-wave forms, triangular forms, and kicks forms.
Among them, the periodic kicks form has been widely used for studying dynamical localization, system cooling, and web-assisted tunneling.\ucite{PhysRevLett.83.4037,PhysRevA.68.063406,PhysRevLett.93.204101,
PhysRevLett.96.160403,PhysRevLett.105.054103,WANG2011,PhysRevE.91.032923,PhysRevA.99.033618,WANG2018,YU2019,PhysRevB.101.064302}
Recently, with the help of the delta-kick collimation technique,\ucite{PhysRevLett.114.143004,PhysRevLett.110.093602,Corgier2020,PhysRevResearch.3.033261,PhysRevLett.127.100401} Bose-Einstein condensates have been well manipulated. In particular, the entanglement in the Bose-Einstein condensate can be remarkably enhanced by using the delta-kick squeezing method.\ucite{PhysRevLett.127.183401}
By a combination of interactions and periodic kicks, multi-particle bound states are generated in a one-dimensional lattice,\ucite{PhysRevB.95.014305,PhysRevB.96.104309} and the nonlinear dynamics and quantum chaos are adequately characterized in the $p$-spin model.\ucite{PhysRevE.103.052212,10692}
More recently, in single-mode bosonic systems, the dynamical blockade is induced by periodic kicks fields in all regimes of nonlinearity strengths.\ucite{PhysRevLett.123.013602}

In this work, we first minutely study the enrichment dynamics of the quantum systems driven by periodic kicks, and show that the kick dynamics can be well described by a time-independent effective Hamiltonian. When properly choosing the parameters of periodic kicks, the system would appear the coherent destruction of coupling phenomenon\ucite{PhysRevLett.67.516,Grifoni1998} in the resonance regime. Particularly, we will see that, the effective coupling strength by periodic kicks can be much larger than the coupling strength of the original system. This result stems
from the instantaneous of kicks, namely zero time duration.
In the same three-level system, we demonstrate that different regimes [e.g., one(two)-photon resonance regime] can be transformed from and to each other by modulating the period of periodic kicks, while other parameters remain unchanged. Finally, we present some applications and physical implementations of periodic kicks.
Note that the relevant results we obtain are quite general to a variety of quantum systems, since our analysis is not restricted to the particular system.

The paper is organized as follows. In Section 2, we introduce the physical model and derive the expression of the effective Hamiltonian in the periodic kicks two-level system. Then, we investigate the effective coupling strength in different kicks, including the frequency kick, the amplitude kick, and the phase kick. Finally, we explore the parameter range in which the effective Hamiltonian is valid.  In Section 3, we generalize the analytical method in a special three-level system, and put forward the sweep method to study the kick dynamics in the general three-level system. In Section 4, we present some applications of periodic kicks in quantum state manipulations.  In Section 5, we demonstrate the physical implementation of periodic kicks in practice. The conclusion is given in Section 6.

\section{Kick dynamics in two-level systems} \label{ii}

\subsection{General formula of the effective Hamiltonian}

Let us first consider a general two-level system driven by a control field, and the Hamiltonian reads ($\hbar=1$)
\begin{eqnarray}  \label{1}
H_1(t)=\Delta_1(t)|2\rangle\langle2|+\Omega_1(t)e^{i\theta_1(t)}|1\rangle\langle2|+\mathrm{H.c.},
\end{eqnarray}
where $\Delta_1(t)$, $\Omega_1(t)$, and $\theta_1(t)$ are called as the detuning, the coupling strength, and the phase, respectively.
In fact, this Hamiltonian is a universal form of describing different quantum systems with the two-level structure. For instance, when employing it to describe the Landau-Zener model, we can set $\theta_1(t)=0$ and $\Delta_1(t)$ represents quasimomentum.\ucite{Bason2011}
When employing it to describe condensed-matter systems, $\Omega_1(t)$ and $\Delta_1(t)$ are related to quasiparticle momenta.\ucite{PhysRevB.93.205415,PhysRevB.94.125423,Gagnon_2016}
When employing it to describe the system of optical driving a two-level atom, $\Delta_1(t)$ and $\Omega_1(t)$ are the detuning and the Rabi frequency, respectively.
Note that the detuning $\Delta_1(t)=\omega_{12}-\omega_c$, where $\omega_{12}$ and $\omega_c$ are the Bohr frequency of the two-level atom and the carrier frequency of the control field.\ucite{scully97}
Remarkably, it is also suited for describing a superconductor qubit, where $\Omega_1(t)$ and $\Delta_1(t)$ are dc and ac flux biases, respectively.\ucite{Chu2004,PhysRevA.79.032301,PhysRevB.96.174518,CPB030307}
Since this Hamiltonian does not impose restrictions on specific systems, we assume that all parameters are adjustable in the following.

When the Hamiltonian in Eq.~(\ref{1}) is time-independent, one readily acquires the evolution operator of this system, and its matrix form in basis $\{|1\rangle, |2\rangle\}$ reads
\begin{eqnarray}   \label{2c}
U_1(t)=e^{-iH_1t}= \left(
                \begin{array}{cc}
                  \cos(E_1t)+i\frac{\Delta_1}{2E_1}\sin(E_1t) & -i\frac{\Omega_1e^{i\theta_1}}{E_1}\sin(E_1t) \\[1ex]
                  -i\frac{\Omega_1e^{-i\theta_1}}{E_1}\sin(E_1t) & \cos(E_1t)-i\frac{\Delta_1}{2E_1}\sin(E_1t) \\
                \end{array}
                 \right),
\end{eqnarray}
where the eigenenergy $E_1=\sqrt{\Omega_1^2+{\Delta_1^2}/{4}}$.
Note that the parentheses are deleted for brevity hereafter when the parameter is time-independent.
In particular, if the system is in the large detuning regime, namely $|\Delta_1|\gg\Omega_1$, the evolution operator approximatively becomes
\begin{eqnarray}
U_1(t)\approx \left(
                \begin{array}{cc}
                  e^{iE_1t} & 0 \\[1ex]
                  0 & e^{-iE_1t} \\
                \end{array}
              \right).
\end{eqnarray}
Therefore, the population evolution of system states is frozen, which is a well-known result.

Here, to see more clearly how the periodic kicks change the system dynamics, we set the system in the large detuning regime. First of all, the Hamiltonian of the system driven by periodic $\delta$-function kicks is
\begin{eqnarray}   \label{2}
H(t)=H_1+H_1'(t),
\end{eqnarray}
where $T$ is the period of kicks, and the form of the kick Hamiltonian $H_1'(t)$ reads
\begin{eqnarray}
H_1'(t)=\sum_{n=-\infty}^{+\infty}\delta(t-nT)\Big[\Delta_1'|2\rangle\langle2| +\Omega_1'e^{i\theta_1'}|1\rangle\langle2|+\mathrm{H.c.}\Big].
\end{eqnarray}

Although the Floquet theory\ucite{PhysRev.138.B979,PhysRevA.7.2203} is a powerful tool to calculate the time-independent effective Hamiltonian of periodic systems,\ucite{Shevchenko2010,Eckardt_2015,Silveri_2015,CPB074205,PhysRevB.94.125101,PhysRevB.94.195108,CPB127308,PhysRevLett.120.243602,
PhysRevA.100.023622,PhysRevLett.122.010407,Yang_2019,CPB023201,PhysRevA.101.022108} it does not work very well in this system due to the singularity of $\delta$-function. In addition, the effective Hamiltonian is hardly obtained by employing the Baker-Campbell-Hausdorff formula\ucite{PhysRevA.68.013820,PhysRevX.4.031027,PhysRevE.91.032923,CPB064204} as well as the rotating wave approximation,\ucite{PhysRevA.75.063414,Silveri_2017,PhysRevB.97.125429,PhysRevB.98.195434,PhysRevLett.120.123204} because we cannot treat the period $T$ as a perturbation in this system due to its arbitrariness.
Here, we put forward an alternative method to achieve the effective Hamiltonian as follows. At first, the evolution operator of this periodic kicks system can be written as a product of two terms: an evolution with the Hamiltonian $H_1$ for the duration $T$ followed by an evolution with the Hamiltonian $H_1'(t)$.
Thus, by making use of Eq.~(\ref{2c}), the evolution operator within one period is expressed as
\begin{eqnarray}  \label{6}
U(T)=e^{-i\mathcal{M}_1}e^{-iH_1T}
    \!=\!\left(
                \begin{array}{cc}
                  f_1\!+\!if_2 &  f_4\!-\!if_3 \\[0.5ex]
                  -\!f_4\!-\!if_3 & f_1\!-\!if_2 \\
                \end{array}
              \right),~~
\end{eqnarray}
where $\mathcal{M}_1=\Delta_1'|2\rangle\langle2| +\Omega_1'e^{i\theta_1'}|1\rangle\langle2|+\mathrm{H.c.}$, and the functions $f_k$ ($k=1,2,3,4$) are
\begin{eqnarray}
f_1\!\!\!&=&\!\!\!\cos E_1'\cos E_1T-\frac{\Delta_1\Delta_1'+4\Omega_1\Omega_1'\cos(\theta_1-\theta_1')}{4E_1E_1'}\sin E_1' \sin E_1T, \\[1ex]
f_2\!\!\!&=&\!\!\!\frac{\Delta_1}{2E_1}\cos E_1'\sin E_1T\!+\!\frac{\Delta_1'}{2E_1'}\sin E_1'\cos E_1T\!+\!\frac{\Omega_1\Omega_1'\sin(\theta_1-\theta_1')}{E_1E_1'}\sin E_1' \sin E_1T, \\[1ex]
f_3\!\!\!&=&\!\!\!\frac{\Omega_1\cos\theta_1}{E_1}\cos \!E_1'\sin\! E_1T\!+\!\frac{\Omega_1'\cos\theta_1'}{E_1'}\sin \!E_1'\cos \!E_1T \!+\!\frac{\Delta_1\Omega_1'\sin\theta_1'\!-\!\Delta_1'\Omega_1\sin\theta_1}{2E_1E_1'}\sin\! E_1'\sin\! E_1T,  \\[1ex]
f_4\!\!\!&=&\!\!\!\frac{\Omega_1\sin\theta_1}{E_1}\cos \!E_1'\sin\! E_1T\!+\!\frac{\Omega_1'\sin\theta_1'}{E_1'}\sin \!E_1'\cos \!E_1T \!+\!\frac{\Delta_1'\Omega_1\cos\theta_1\!-\!\Delta_1\Omega_1'\cos\theta_1'}{2E_1E_1'}\sin \!E_1'\sin\! E_1T,
\end{eqnarray}
and the eigenenergy $E_1'=\sqrt{\Omega_1'^2+{\Delta_1'^2}/{4}}$.
Next, we presuppose the time-independent effective Hamiltonian has a general form
\begin{eqnarray}  \label{3}
H_{\mathrm{eff}}=\Delta_{\mathrm{eff}}|2\rangle\langle2|+\Omega_{\mathrm{eff}}e^{i\theta_{\mathrm{eff}}}|1\rangle\langle2|+\mathrm{H.c.},
\end{eqnarray}
where the unknown coefficients $\Delta_{\mathrm{eff}}$, $\Omega_{\mathrm{eff}}$, and $\theta_{\mathrm{eff}}$ are yielded by inversely solving the equation $U(T)=\exp(-iH_{\mathrm{eff}}T)$.\ucite{Eastham73} After some simple calculations, the expressions are given by
\begin{eqnarray} \label{6b}
\Delta_{\mathrm{eff}}=\frac{2f_2} {T\sqrt{1-f_1^2}}\arccos|f_1|, ~~~~~
\Omega_{\mathrm{eff}}=\sqrt{\frac{f_3^2+f_4^2} {T^2(1-f_1^2)}}\arccos|f_1|, ~~~~~
\theta_{\mathrm{eff}}=\arctan\frac{f_4}{f_3}.
\end{eqnarray}
As a result, we achieve the general formula of the effective Hamiltonian for this periodic kicks system.

\begin{figure}
\begin{center}
\includegraphics[width=6in]{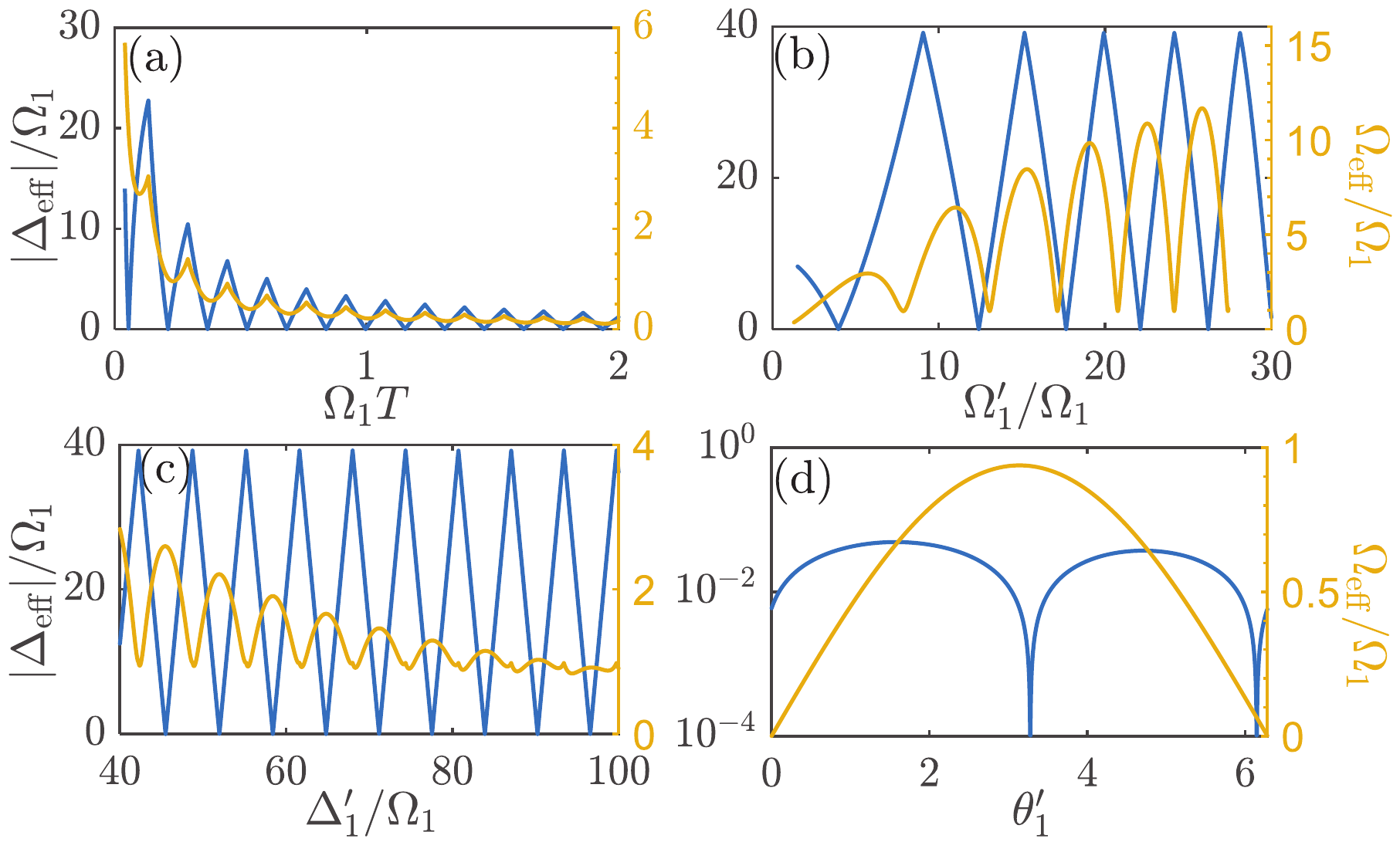}\\[5pt]  
\caption{ (color online) Effective detuning $\Delta_{\mathrm{eff}}$ (the blue curves) and coupling strength $\Omega_{\mathrm{eff}}$ (the yellow curves) in different periodic kicks, where the parameters are given in Table~\ref{bt1}.}\label{fig:01}
\end{center}
\end{figure}

Compared to the original system described by the Hamiltonian in Eq.~(\ref{1}), we observe from Eq.~(\ref{3}) that
the kick dynamics of the two-level system is now controlled by the effective detuning $\Delta_{\mathrm{eff}}$, the effective coupling strength $\Omega_{\mathrm{eff}}$ and the effective phase $\theta_{\mathrm{eff}}$, and those effective parameters depend on the parameters $\{T,\Omega_1',\Delta_1',\theta_1'\}$ of the periodic kicks.
Figures~\ref{fig:01}(a)-\ref{fig:01}(d) are plotted the effective detuning $\Delta_{\mathrm{eff}}$ and the effective coupling strength $\Omega_{\mathrm{eff}}$ as a function of the parameters $T$, $\Omega_1'$, $\Delta_1'$, and $\theta_1'$, respectively.
The results demonstrate that $\Delta_{\mathrm{eff}}$ and $\Omega_{\mathrm{eff}}$ emerge distinct behaviors and both of them are variable when changing different parameters of periodic kicks.
Figure~\ref{fig:01} also demonstrates that even though the original system is in the large detuning regime, the populations of quantum states still evolves if $\Delta_{\mathrm{eff}}=0$ and $\Omega_{\mathrm{eff}}\neq0$.

\renewcommand\arraystretch{1.2}
\begin{table}[htbp]
	\centering
	\caption{Values of physical parameters (in units of $\Omega_1$) in the two-level system.}
	\label{bt1}
	\begin{tabular}{cccccccc}
		\hline
           ~~&~~ {$\Delta_1$} ~~&~~{$\theta_1$} ~~&~~{$T$} ~~&~~ {${\Omega_1'}$}~~&~~ {$\Delta_1'$} ~~&~~ {$\theta_1'$}  \\
		\hline
        {Fig.~\ref{fig:01}(a)} &  40  &~ 0 & --  & 6   & 40   &~ 0    \\
        {Fig.~\ref{fig:01}(b)} &  40  &~ 0 & 0.08  & --   & 40   &~ 0    \\
        {Fig.~\ref{fig:01}(c)} &  40  &~ 0 & 0.08  & 6   & --   &~ 0    \\
        {Fig.~\ref{fig:01}(d)} &  40  &~ 0 &~ 0.0982   & 1   & 40   &~ --    \\
        {Fig.~\ref{fig:02}(a)} &  100  &~ 0 &~ 0.0628    & 1   & 56.5133   &~ 0    \\
        {Fig.~\ref{fig:02}(b)} &  100  &~ 0 &~ --   & 1   & --   &~ 0    \\
        {Fig.~\ref{fig:02}(c)} &  40  &~ 0 &~ --   & --   & 40   &~ 0    \\
        {Fig.~\ref{fig:21}} &  40  &~ 0 &~ 0.05    & --   & --   &~ --    \\
        {Fig.~\ref{fig:03s}}(a) & 20   &~ 0  &~ 0.1359    & 5   &20    &~0     \\
        {Fig.~\ref{fig:03s}}(b) & 35   &~ 0 &~ 0.0506    & 4   &35    &~0    \\
		\hline	
	\end{tabular}
\end{table}

\subsection{Effective coupling strength in the resonance regime}   \label{3.1}

In this subsection, we focus on the effective coupling strength when the periodic kicks system is in the resonance regime, i.e., $\Delta_{\mathrm{eff}}=0$.
It is easily observed from Eq.~(\ref{6b}) that the zero-points of the effective detuning $\Delta_{\mathrm{eff}}$ are only determined by the roots of the equation $f_2=0$. In Appendix~A, we demonstrate that not all parameters $\{T,\Omega_1',\Delta_1',\theta_1'\}$ can be individually modulated to guarantee the periodic kicks system is in the resonance regime, i.e., $\Delta_{\mathrm{eff}}=0$. For example, when $\theta_1'$ is variable, the zeros of the function $f_2$ does not always exist for arbitrary values of $\{T, \Omega_1',\Delta_1'\}$.

Since the parameters $\{T,\Omega_1',\Delta_1',\theta_1'\}$ can be individually controllable, there are several adjustment styles to be employed in periodic kicks. For simplicity, we first study the style in which we only periodically kick on the detunings $\{\Delta_1, \Delta_1'\}$, while the other parameters remain unchanged (namely $\Omega_1'=\Omega_1$ and $\theta_1'=\theta_1$). In particular, the period $T$ of periodic kicks are determined by solving the equation $\Delta_{\mathrm{eff}}=0$. According to Appendix~A, the expression of the period is given by
\begin{eqnarray}
T=\frac{-\phi_1+n\pi}{E_1}, ~~~n\in\mathbb{N},
\end{eqnarray}
where $\mathbb{N}$ represents integers, and
\begin{eqnarray}
\phi_1=\arctan\frac{\Delta_1'E_1\sin E_1'}{\Delta_1E_1'\cos E_1'+2\Omega_1\Omega_1'\sin(\theta_1-\theta_1')\sin E_1'}.
\end{eqnarray}
Then, the expression of the effective coupling strength $\Omega_{\mathrm{eff}}$ is reduced as
\begin{eqnarray}
\Omega_{\mathrm{eff}}=\frac{\arccos f_1}{T}.
\end{eqnarray}
We call this adjustment style as the frequency kick for briefness hereafter.

\begin{figure}
\begin{center}
\includegraphics[width=5in]{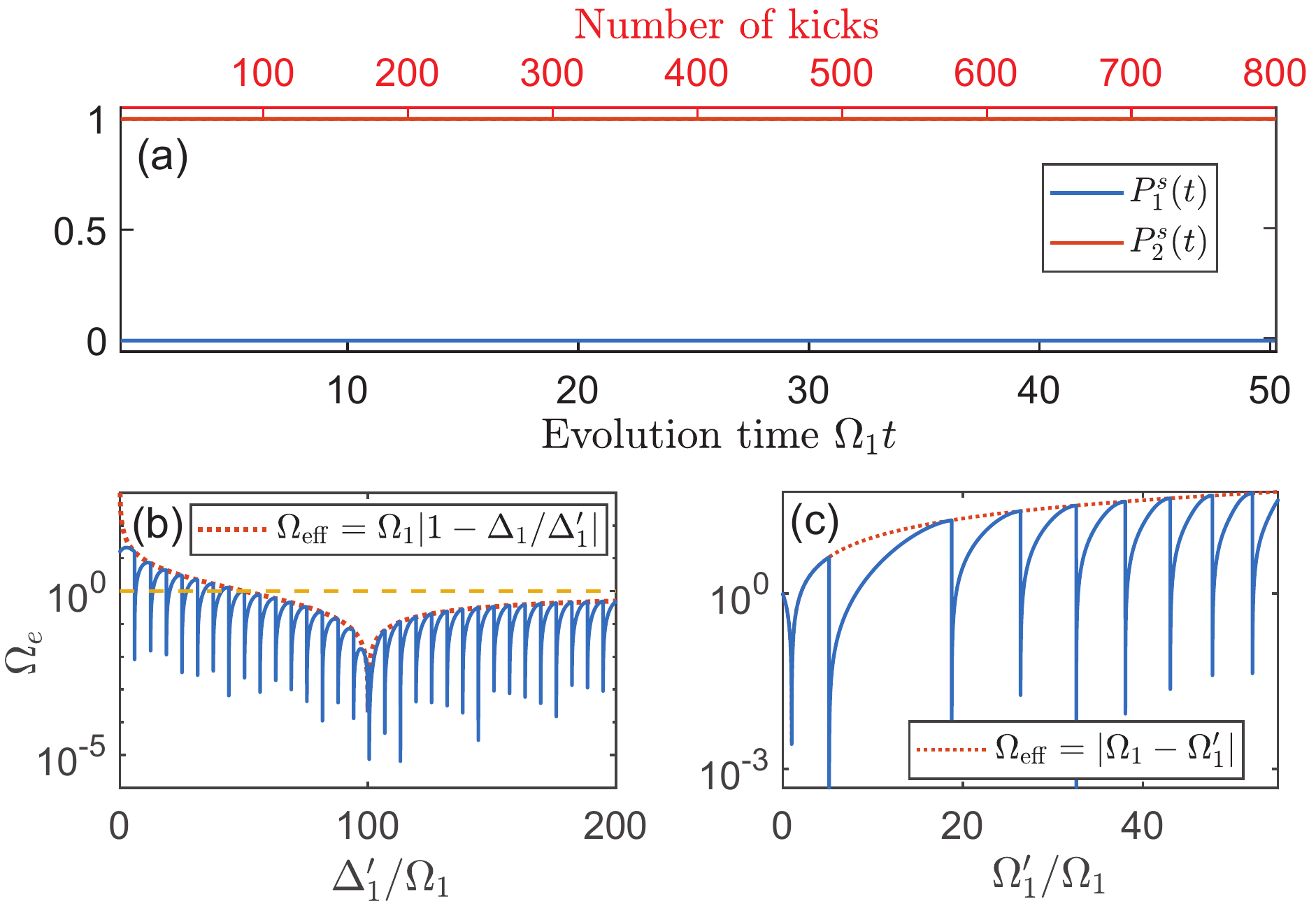}  
\caption{ (color online) (a) Time evolution of the state $|j\rangle$ ($j=1,2$) in the frequency kick, where the physical parameters satisfy $E_1'/\Omega_1=9\pi$.
(b) $\Omega_{\mathrm{eff}}$ vs $\Delta_1'$ in the frequency kick.
(c) $\Omega_{\mathrm{eff}}$ vs $\Omega_1'$ in the amplitude kick. The parameters are given in Table~\ref{bt1}. }\label{fig:02}
\end{center}
\end{figure}

When the detuning $\Delta_1'$ is tuned to satisfy
\begin{eqnarray}
E_1'=\sqrt{\Omega_1'^2+{\Delta_1'^2}/{4}}=m\pi, ~~m\in\mathbb{N},
\end{eqnarray}
we find that the effective coupling between the states $|1\rangle$ and $|2\rangle$ would vanish, i.e., $\Omega_{\mathrm{eff}}=0$; see Appendix~B for details. In fact, this phenomenon is known as the coherent destruction of coupling,\ucite{PhysRevLett.67.516,Grifoni1998} where the system dynamics is frozen even in the resonance regime ($\Delta_{\mathrm{eff}}=0$). We plot in Fig.~\ref{fig:02}(a) the dynamical evolution of the two-level system driven by the frequency kick, where
\begin{eqnarray}
P_j^{s}(t)=|\langle j|\psi^s(t)\rangle|^2,~~~j=1,2,
\end{eqnarray}
represents the population of the state $|j\rangle$ at the evolution time $t$, and $|\psi^s(t)\rangle$ represents the system state governed by the Hamiltonian in Eq.~(\ref{2}).
It is shown that the population almost keeps unchanged during the whole evolution process. Physically, this phenomenon results from the destructive quantum interference.\ucite{PhysRevLett.67.516,Grifoni1998}
Furthermore, we can see that the coherent destruction of coupling phenomenon is independent on the parameters of the original system and only decided by the parameters of periodic kicks.

Although $\Omega_{\mathrm{eff}}=0$ at $E_1'=m\pi$ in the frequency kick, it is quite difference at the limitation of $E_1'\rightarrow(m\pi)^{-}$. As shown in Appendix~B, the effective coupling strength at this limitation reads
\begin{eqnarray} \label{4}
\lim\limits_{{E_1'\rightarrow(m\pi)^{-}}}\Omega_{\mathrm{eff}}=\Omega_1|1-\frac{\Delta_1}{\Delta_1'}|.
\end{eqnarray}
Remarkably, the effective coupling strength $\Omega_{\mathrm{eff}}$ increases with the increase of the detuning $\Delta_1$.
Particularly, $\Omega_{\mathrm{eff}}$ would be infinity when $\Delta_1\rightarrow\infty$, even though the coupling strength $\Omega_1$ of the original system is finite.
In fact, the enlargement of the effective coupling strength stems from the trick that the kicks are instantaneous, i.e., zero time duration.
We plot in Fig.~\ref{fig:02}(b) the effective coupling strength $\Omega_{\mathrm{eff}}$ as a function of the detuning $\Delta_1'$ in the frequency kick.
For comparison, we also plot the curve $\Omega_{\mathrm{eff}}=\Omega_1|1-{\Delta_1}/{\Delta_1'}|$ (the dot curve) in Fig.~\ref{fig:02}(b), which demonstrate that the analytical expression is in agreement with the numerical results.

When we only periodically kick on the coupling strength $\{\Omega_1, \Omega_1'\}$ and keep the other parameters unchanged (we call this adjustment style as the amplitude kick), the effective coupling strength at the limitation of $E_1'\rightarrow(m\pi)^{-}$ reads
\begin{eqnarray} \label{5}
\lim\limits_{{E_1'\rightarrow(m\pi)^{-}}}\Omega_{\mathrm{eff}}=|\Omega_1-\Omega_1'|.
\end{eqnarray}
In this situation, the effective coupling strength $\Omega_{\mathrm{eff}}$ is always smaller than the coupling strength of the original system, namely $\Omega_{\mathrm{eff}}<\max\{\Omega_1,\Omega_1'\}$. Figure~\ref{fig:02}(c) shows the effective coupling strength $\Omega_{\mathrm{eff}}$ as a function of the coupling strength $\Omega_1'$ in the amplitude kick.
For comparison, we also plot the curve $\Omega_{\mathrm{eff}}=|\Omega_1-\Omega_1'|$ (the dot curve) in Fig.~\ref{fig:02}(c). One can observe in Figs.~\ref{fig:02}(b)-\ref{fig:02}(c) that $\Omega_{\mathrm{eff}}$ exhibits periodic variations over parameters and $\Omega_{\mathrm{eff}}=0$ at the points $E_1'=m\pi$.

Finally, when we only periodically kick on the phase $\{\theta_1, \theta_1'\}$ and keep the other parameters unchanged (we call this adjustment style as the phase kick), in the large detuning regime, the effective coupling strength at the limitation of $E_1'\rightarrow(m\pi)^{-}$ approximatively reads
\begin{eqnarray} \label{5}
\lim\limits_{{E_1'\rightarrow(m\pi)^{-}}}\Omega_{\mathrm{eff}}\approx\Omega_1|1-e^{i(\theta_1-\theta_1')}|.
\end{eqnarray}
In this situation, the effective coupling strength $\Omega_{\mathrm{eff}}$ can also be much larger than the coupling strength $\Omega_1$ of the original system by the appropriate value of $\theta_1'$.

\subsection{Validity of the effective Hamiltonian}

In the derivation of the expression of the effective Hamiltonian~(\ref{3}), we do not employ any approximations.
This means that the effective Hamiltonian~(\ref{3}) can exactly describe the kick dynamics of the system for all parameter range at the evolution time $t=nT$. The natural question is that whether the effective Hamiltonian~(\ref{3}) is still valid for describing the kick dynamics when $t\neq nT$.
In this subsection, we focus on this issue.

To quantify the validity of the effective Hamiltonian~(\ref{3}), we adopt the following definition
\begin{eqnarray}
P_2^{d}\!\!\!&=&\!\!\!\max\Big\{|P_2^{s}(t)-P_2^{e}(t)|\Big\}, \nonumber\\[0.5ex]
\!\!\!&=&\!\!\!\max\Big\{|\langle 2|\psi^s(t)\rangle|^2-|\langle 2|\psi^e(t)\rangle|^2|\Big\},~~~
\end{eqnarray}
where $|\psi^s(t)\rangle$ and $|\psi^e(t)\rangle$ are the evolution state of the system governed by the system Hamiltonian~(\ref{2}) and the effective Hamiltonian~(\ref{3}), respectively. Obviously, $P_2^{d}$ quantifies the maximum population deviation of the state $|2\rangle$ during the dynamical evolution described by this two Hamiltonians.
Note that a high value of $P_2^{d}$ means that the effective Hamiltonian~(\ref{3}) cannot justify the kick dynamics of the system. On the contrary, the effective Hamiltonian~(\ref{3}) would work well if $P_2^{d}$ is low.

\begin{figure}
\begin{center}
\includegraphics[width=5in]{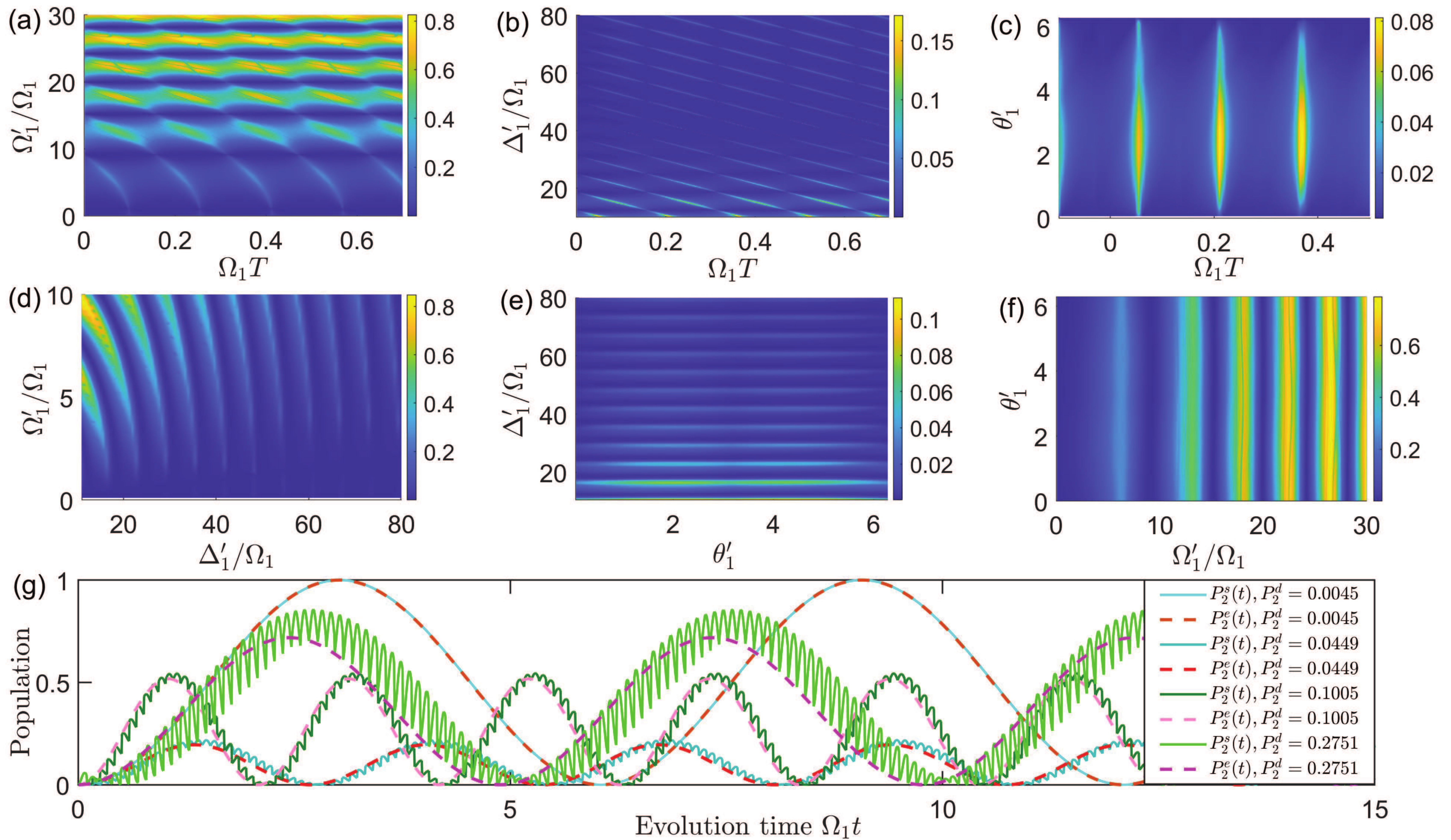}\\[5pt]  
\caption{ (color online) (a) $P_2^d$ vs $\Omega_1'$ and $T$. (b) $P_2^d$ vs $\Delta_1'$ and $T$. (c) $P_2^d$ vs $\theta_1'$ and $T$. (d) $P_2^d$ vs $\Omega_1'$ and $\Delta_1'$. (e) $P_2^d$ vs $\Delta_1'$ and $\theta_1'$. (f) $P_2^d$ vs $\theta_1'$ and $\Omega_1'$. The parameters of the original system are given in Table~\ref{bt1}, and the remaining parameters of periodic kicks are the same as those of the original system. (g) Time evolution of the state $|2\rangle$ governed by the system Hamiltonian~(\ref{2}) (the solid curves) and the effective Hamiltonian~(\ref{3}) (the dash curves) by choosing different values of $P_2^d$, where the initial state is $|1\rangle$.}\label{fig:21}
\end{center}
\end{figure}

Figures~\ref{fig:21}(a)-\ref{fig:21}(f) demonstrate $P_2^{d}$ versus different parameters of periodic kicks.
We can see that $P_2^{d}$ is low in most parameter regions and presents regular change with the parameters of periodic kicks. Note that $P_2^{d}$ would be relatively high in some regions. The reason is that the periodic kicks system is usually in the (near-)resonance regime (i.e., $\Delta_{\mathrm{eff}}\approx0$) in those regions.
Furthermore, Figs.~\ref{fig:21}(b), \ref{fig:21}(d), and \ref{fig:21}(e) indicate that $P_2^d$ decreases with the increase of the detuning $\Delta_1'$, which means the effective Hamiltonian~(\ref{3}) gives a very good performance in describing the kick dynamics when the original system is in the large detuning regime.
We also find from Figs.~\ref{fig:21}(a), \ref{fig:21}(d), and \ref{fig:21}(f) that $P_2^d$ is high if the value of $\Omega_1'$ is comparable to the value of $\Delta_1'$, i.e., $\Omega_1'\gtrsim\Delta_1'$.
Physically, it originates from the fact that the system states quite readily evolve when $\Omega_1'\gtrsim\Delta_1'$ in the kick Hamiltonian $H_1'(t)$.
Finally, Figs. \ref{fig:21}(c), \ref{fig:21}(e), and \ref{fig:21}(f) demonstrate that $P_2^d$ is little related to the phase $\theta_1'$, since the phase $\theta_1'$ cannot change the energy gap of two-level system (cf. the eigenenergy $E_1'=\sqrt{\Omega_1'^2+{\Delta_1'^2}/{4}}$).

To see the different dynamics described by the system Hamiltonian~(\ref{2}) and the effective Hamiltonian~(\ref{3}) more clearly, we present in Fig.~\ref{fig:21}(g) some specific examples of the dynamical evolution by choosing different values of $P_2^{d}$.
The results show that the kick dynamics of the system are well described by the effective Hamiltonian~(\ref{3}) in the long timescale, even though $P_2^d$ is as high as 0.2751. Primarily, it becomes more accurate when $P_2^d$ is low.

\section{Kick dynamics in three-level systems} \label{iv}

\subsection{General description of three-level systems}

In this section, we study the dynamics of the three-level system driven by periodic kicks. For a three-level system interacting with two control fields, the general form of the Hamiltonian is given by
\begin{equation}  \label{8}
H_1=\sum_{l=1}^{2}\Delta_l|l\!+\!1\rangle\langle l\!+\!1|+\Omega_le^{i\theta_l}|l\rangle\langle l\!+\!1|+\mathrm{H.c.},
\end{equation}
where $\Delta_l$, $\Omega_l$, and $\theta_l$ ($l=1,2$) also represent the detuning, the coupling strength, and the phase, respectively.
{Note that this three-level model is easily found in various quantum systems, including atomic systems,\ucite{scully97} superconducting circuits,\ucite{rmp623} and trapped ions,\ucite{rmp281} etc.
For example, the Rydberg state excitation from the ground state is actually a three-level model,\ucite{prl113002} because it is hard to obtain a laser field directly driving the Rydberg state from the ground state, and thus one requires two laser fields to get this process through an intermediate state.
Moreover, some complex quantum systems can also be simplified as the three-level model.\ucite{scpma59,pra062338,pra042328,njp24}}

It is not hard to obtain the eigenenergies $E_n$ and eigenstates $|E_n\rangle$ ($n=1,2,3$) of the Hamiltonian in Eq.~(\ref{8}), which are
\begin{eqnarray}\label{9a}
E_n\!\!\!&=&\!\!\!\frac{\Delta_1+\Delta_2}{3}+(-\frac{4p}{3})^{\frac{1}{2}}\cos(u+\vartheta_n),\\
|E_n\rangle\!\!\!&=&\!\!\!C_n\Big[\Omega_1e^{i\theta_1}(E_n-\Delta_2)|1\rangle+E_n(E_n-\Delta_2)|2\rangle +\Omega_2e^{-i\theta_2}E_n|3\rangle\Big],  \label{9c}
\end{eqnarray}
where $C_n$ are normalization constants, $\vartheta_1=0$, $\vartheta_2={2\pi}/{3}$, $\vartheta_3={4\pi}/{3}$, and
\begin{eqnarray}
u\!\!\!&=&\!\!\!\frac{1}{3}\arccos\left[-\frac{q}{2}(-\frac{p}{3})^{-\frac{3}{2}}\right], \\[1.0ex]
p\!\!\!&=&\!\!\!\Delta_1\Delta_2-\Omega_1^2-\Omega_2^2-\frac{(\Delta_1+\Delta_2)^2}{3}, \\[1.0ex]
q\!\!\!&=&\!\!\!\Delta_2\Omega_1^2+\frac{(\Delta_1\!+\!\Delta_2)(\Delta_1\Delta_2\!-\!\Omega_1^2\!-\!\Omega_2^2)}{3} -\frac{2(\Delta_1\!+\!\Delta_2)^3}{27}.
\end{eqnarray}
With the help of Eqs.~(\ref{9a})-(\ref{9c}), the analytical expression of the evolution operator is available in principle, but its complicated form hinders one from further figuring out the dynamical behavior of this three-level system.

Qualitatively, if $\Delta_1=\Delta_2=0$, i.e., the system is in the resonance regime, the $|1\rangle\leftrightarrow|j\rangle$ ($j=2,3$) transition can occur during the evolution process.
If $\Delta_1=0$ and $|\Delta_2|\gg\Omega_l$, i.e., the system is in the one-photon resonance regime, the $|1\rangle\leftrightarrow|2\rangle$ transition can occur during the evolution process. If $\Delta_2=0$ and $|\Delta_1|\gg\Omega_l$, i.e., the system is in the two-photon resonance regime, the $|1\rangle\leftrightarrow|3\rangle$ transition can occur during the evolution process.
However, if $|\Delta_l|\gg\Omega_m$ ($l,m=1,2$), i.e., the system is in the large detuning regime, the dynamics are frozen so that there are no transitions between quantum states.
Generally speaking, it is difficult to make the same system transfer from one regime to another regime, since the frequencies of control fields, such as the laser field, are usually fixed and rigid to adjust.\ucite{scully97} As a result, the system is always modulated in a specific regime, either the resonance regime, the one-photon regime, or the two-photon regime. In the following, we demonstrate that this situation can be changed by periodic kicks.

When the three-level system is driven by periodic kicks, the Hamiltonian has the form
\begin{eqnarray}      \label{16a}
H(t)=H_1+H_1'(t),
\end{eqnarray}
where $T$ is the period of periodic kicks, and the kick Hamiltonian $H_1'(t)$ is
\begin{equation}
H_1'(t)=\sum_{n=-\infty}^{+\infty}\delta(t-nT)\left[\sum_{l=1}^{2}\Delta_l'|l\!+\!1\rangle\langle l\!+\!1|+\Omega_l'e^{i\theta_l'}|l\rangle\langle l\!+\!1|+\mathrm{H.c.}\right].
\end{equation}

\subsection{The special three-level system}

When the three-level system is in the resonance regime ($\Delta_1=\Delta_2=0$) or the two-photon resonance regime ($\Delta_2=0$), the solutions of Eqs.~(\ref{9a})-(\ref{9c}) degenerate into the square root form so that the analytical expression of the evolution operator is easily obtained. Therefore, those regimes have been extensively studied. Nevertheless, those regimes are beyond our consideration. Here, we investigate the original three-level system in the large detuning regime, i.e., $|\Delta_l|\gg\Omega_m$ ($l,m=1,2$). First, we study a special case, where the parameters are chosen as $\Delta_2=2\Delta_1$, $\Omega_2=\Omega_1$, and $\theta_2=\theta_1$.
Under proper periodic kicks, this three-level system can be driven in the resonance regime, i.e., $\Delta_{\mathrm{eff},1}=\Delta_{\mathrm{eff},2}=0$; see Appendix~C for details.
To this end, we only need to properly modulate the period $T$ of periodic kicks, and the expression is given by
\begin{eqnarray}
T=\frac{1}{E_1}\left({\arcsin\frac{-a_{23}}{\sqrt{a_{21}^2+a_{22}^2}}-\arctan\frac{a_{22}}{a_{21}}+2n\pi}\right),
\end{eqnarray}
or
\begin{eqnarray}
T=\frac{1}{E_1}\left[{\arcsin\frac{-a_{23}}{\sqrt{a_{21}^2+a_{22}^2}}-\arctan\frac{a_{22}}{a_{21}}+(2n+1)\pi}\right],
\end{eqnarray}
where $n\in\mathbb{N}$. In particular, the effective coupling strength and phase read (see Appendix C for details)
\begin{eqnarray}
\Omega_{\mathrm{eff}}=\frac{\sqrt{2}}{T}\arccos\frac{\sqrt{g_1(T)}}{E_1E_1'},~~~~~ \theta_{\mathrm{eff}}=\arctan\frac{g_3(T)}{g_4(T)},
\end{eqnarray}
where $E_1=\sqrt{\Delta_1^2+2\Omega_1^2}$ and $E_1'=\sqrt{\Delta_1'^2+2\Omega_1'^2}$.
As a result, the effective Hamiltonian of the periodic three-level system reads
\begin{eqnarray}  \label{19a}
H_{\mathrm{eff}}=\Omega_{\mathrm{eff}}e^{i\theta_{\mathrm{eff}}}(|1\rangle\langle 2|+|2\rangle\langle 3|)+\mathrm{H.c.}
\end{eqnarray}

\begin{figure}
\begin{center}
\includegraphics[width=5in]{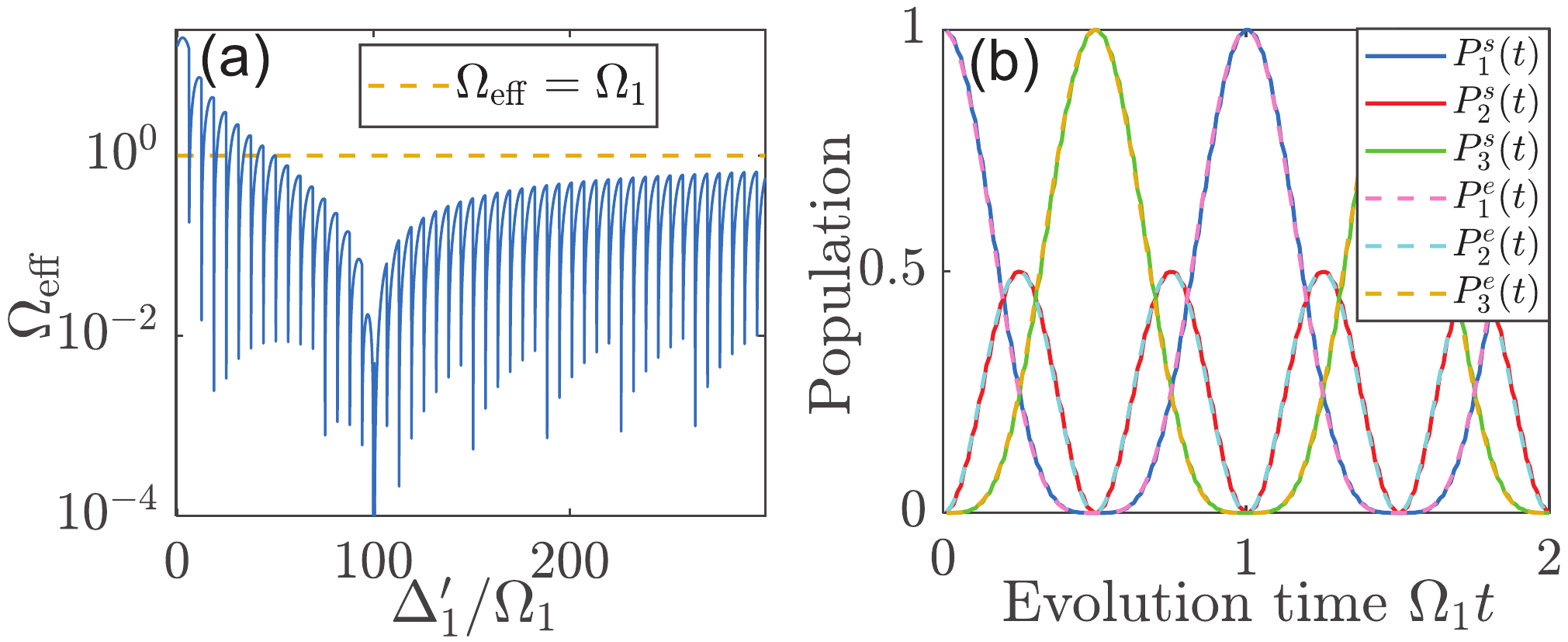}\\[5pt]  
\caption{  (color online) (a) $\Omega_{\mathrm{eff}}$ vs $\Delta_1'$ in the frequency kick. (b) Time evolution of states $|j\rangle$ ($j=1,2,3$) in the three-level system. The parameters are given in Table~\ref{bt2}.}\label{fig:04}
\end{center}
\end{figure}

Figure~\ref{fig:04}(a) demonstrates the effective coupling strength $\Omega_{\mathrm{eff}}$ as a function of $\Delta_1'$ in the frequency kick. Remarkably, it also exists the coherent destruction of coupling phenomenon in the three-level system, and the effective coupling strength can be much larger than the original coupling strength for some regions in the frequency kick.
As a concrete example, Fig.~\ref{fig:04}(b) demonstrates the population evolution in this three-level system, where the solid and dashed curves represent the dynamical evolution governed by the Hamiltonian in Eq.~(\ref{16a}) and the effective Hamiltonian in Eq.~(\ref{19a}), respectively. It is obvious that both the solid and dashed curves are well consistent with each other, directly testifying that the kick dynamics of the three-level system can be well described by the effective Hamiltonian in Eq.~(\ref{19a}).

\renewcommand\arraystretch{1.2}
\begin{table}[htbp]
	\centering
	\caption{Values of physical parameters (in units of $\Omega_1$) in the three-level system, where $\theta_1=\theta_2=0$.}
	\label{bt2}
	\begin{tabular}{ccccccccccc}
		\hline
           & {$\Omega_2$}& {$\Delta_1$} & {$\Delta_2$} & {$T$} & {${\Omega_1'}$} & {${\Omega_2'}$} & {$\Delta_1'$} & {$\Delta_2'$} & {$\theta_1'$} & {$\theta_2'$}  \\
		\hline
        {Fig.~\ref{fig:04}(a)} & 1 & 100 &  200   & --  & 1   & 1   & -- & --   & 0   & 0    \\
        {Fig.~\ref{fig:04}(b)} & 1 & 100 &  200   & --  & 1   & 1   & 18 & 32   & 0   & 0    \\
        {Figs.~\ref{fig:S4}(a)-(c)}~ & 2 & 60 &  40  & --  & --   & 2   & 60 & 40   & 0   & 0    \\
        {Figs.~\ref{fig:S4}(d)-(e)} & 2 & 60 &  40   & --  & 1.5   & 2   & 60 & 40   & 0   & 0    \\
        {Fig.~\ref{fig:S4}(f)} & 2 & 60 &  40   & 0.0424  & 1.5   & 2   & 60 & 40   & 0   & 0    \\
        {Fig.~\ref{fig:S4}(g)} & 2 & 60 &  40  & 0.104  & 1.5   & 2   & 60 & 40   & 0   & 0    \\
        {Figs.~\ref{fig:S6}(a)-(c)} & 2 & 60 &  40  & --  & 1   & 2   & 60 & 40   & --   & 0    \\
        {Figs.~\ref{fig:S6}(d)-(e)} & 2 & 60 &  40   & --  & 1   & 2   & 60 & 40   & $\pi$   & 0    \\
        {Fig.~\ref{fig:S6}(f)} & 2 & 60 &  40   & 0.04318  & 1   & 2   & 60 & 40   & $\pi$   & 0    \\
        {Fig.~\ref{fig:S6}(g)} & 2 & 60 &  40   & 0.1046  & 1   & 2   & 60 & 40   & $\pi$   & 0    \\
		\hline
	\end{tabular}
\end{table}

\subsection{The general three-level system}

When $\Delta_2\neq2\Delta_1$, $\Omega_2\neq\Omega_1$, or $\theta_2\neq\theta_1$, the analytical expression of the evolution operator $U(t)$, as well as the period $T$ of periodic kicks, are tedious and intricate.
The most direct way of achieving the period $T$ is to adopt the sweep method by numerical simulations. The specific process is as follows.

At first, the initial state of the parameter-given three-level system is set to be in the ground state $|1\rangle$.
Then, we alter the value of the period $T$ from 0 to a upper bound value. For each value of the period $T$, we numerically solve the Schr\"{o}dinger equation:
\begin{eqnarray} \label{20}
|\dot{\psi}^s(t)\rangle=-iH(t)|\psi^s(t)\rangle,~~~~~|\psi^s(0)\rangle=|1\rangle,
\end{eqnarray}
where the Hamiltonian $H(t)$ is given by Eq.~(\ref{16a}) and $|\psi^s(t)\rangle$ is the system state at the evolution time $t$.
According to Eq.~(\ref{20}), we would learn about the population evolution $P_j^{s}(t)=|\langle j|\psi^s(t)|^2$ ($j=1,2,3$) of each state in this three-level system. Then, we easily yield the minimum population $P_1^{\mathrm{min}}$ of the ground state $|1\rangle$ and the maximum population $P_j^{\mathrm{max}}$ of the excited state $|j\rangle$ ($j=2,3$) during the whole evolution process, where the definitions are
\begin{eqnarray}
P_1^{\mathrm{min}}\!\!\!&=&\!\!\!\min\{P_1^s(t)\},   \\[0.5ex]
P_j^{\mathrm{max}}\!\!\!&=&\!\!\!\max\{P_j^s(t)\},~~j=2,3.
\end{eqnarray}
For a particular value of the period $T$, if $P_2^{\mathrm{max}}$ ($P_3^{\mathrm{max}}$) nearly approaches 1, it means that there exists resonance coupling between the ground state $|1\rangle$ and the excited state $|2\rangle$ ($|3\rangle$), and this value is what we search for to make the periodic kicks system in the one(two)-photon resonance regime.
If $P_1^{\mathrm{min}}$ does not approach 0 for some values of the period $T$, it means that the periodic system cannot be driven into the resonance regime.
In the following, we illustrate some examples to explain how this sweep method works in detail.
Since periodic kicks have several adjustable parameters, we also divide those situations into three types: the amplitude kick, the phase kick, and the frequency kick.

\begin{figure}
\begin{center}
\includegraphics[width=6in]{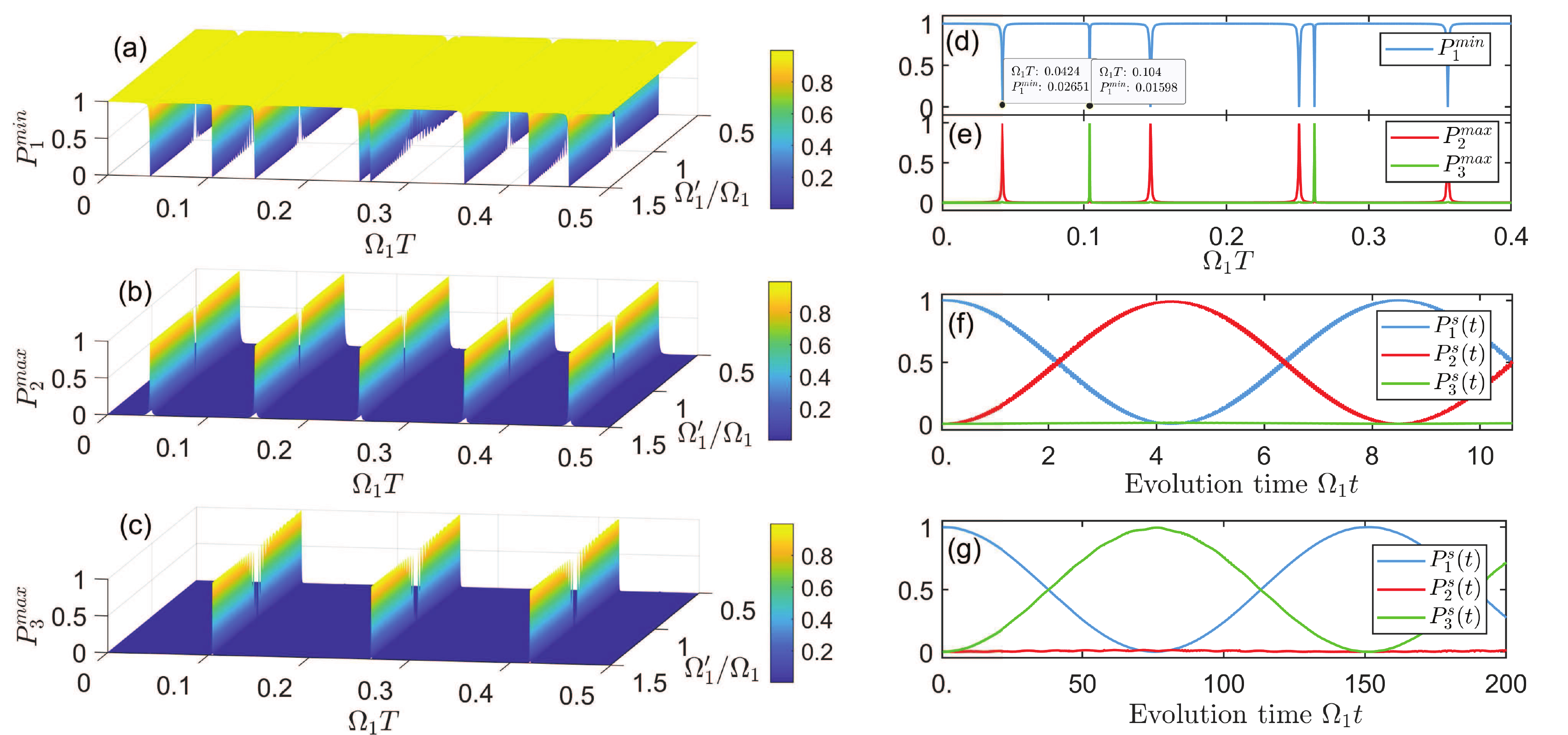}\\[5pt]  
\caption{  (color online) (a) $P_1^{\mathrm{min}}$, (b) $P_2^{\mathrm{max}}$, and (c) $P_3^{\mathrm{max}}$ vs $T$ and $\Omega_1'$, where the parameters are given in Table~\ref{bt2}. (d) $P_1^{\mathrm{min}}$ and (e) $P_j^{\mathrm{max}}$ $(j=2,3)$ vs $T$ by setting $\Omega_1'/\Omega_1=1.5$ in panels (a)-(c). (f) Population evolution of the state $|j\rangle$ $(j=1,2,3)$, where the system is driven into the one-photon resonance regime. (g) Population evolution of the state $|j\rangle$ $(j=1,2,3)$, where the system is driven into the two-photon resonance regime.  }\label{fig:S4}
\end{center}
\end{figure}

For the amplitude kick, we can modulate $\Omega_1'$ or $\Omega_2'$ while the other parameters remain unchanged, namely $\Delta_l'=\Delta_l$ and $\theta_l'=\theta_l$ ($l=1,2$).
We illustrate this situation by periodically kicking the coupling strength $\Omega_1'$.
In Figs.~\ref{fig:S4}(a)-\ref{fig:S4}(c), we plot the minimum population $P_1^{\mathrm{min}}$, the maximum population $P_2^{\mathrm{max}}$, and the maximum population $P_3^{\mathrm{max}}$ as a function of the period $T$ and the coupling strength $\Omega_1'$.
It shows in Figs.~\ref{fig:S4}(a)-\ref{fig:S4}(c) that the three-level system is driven into distinct regimes at specific period $T$ with regular intervals.
Furthermore, the value of the coupling strength $\Omega_1'$ hardly affects the dynamical behavior, since the original system is in the large detuning regime. This result directly demonstrates that the amplitude kick is robust against the amplitude noise of control fields.
To see more clearly, we plot in Figs.~\ref{fig:S4}(d)-\ref{fig:S4}(e) the two-dimensional version of Figs.~\ref{fig:S4}(a)-\ref{fig:S4}(c) by setting $\Omega_1'/\Omega_1=1.5$.
In Fig.~\ref{fig:S4}(d), we easily seek out the value of the period $T$ to make the system in distinct resonance regimes, namely those points making the value of $P_1^{\mathrm{min}}$ nearly approaching zero. And two examples that the system is in the one-photon ($\Delta_{\mathrm{eff},1}=0$) and two-photon ($\Delta_{ \mathrm{eff},2}=0$) resonance regimes are respectively plotted in Figs.~\ref{fig:S4}(f)-\ref{fig:S4}(g).
The results show that the Rabi-like oscillation only appears between $|1\rangle$ and $|2\rangle$ ($|3\rangle$) by periodic kicks when the original three-level system is driven into the one(two)-photon resonance regime.

\begin{figure}
\begin{center}
\includegraphics[width=6in]{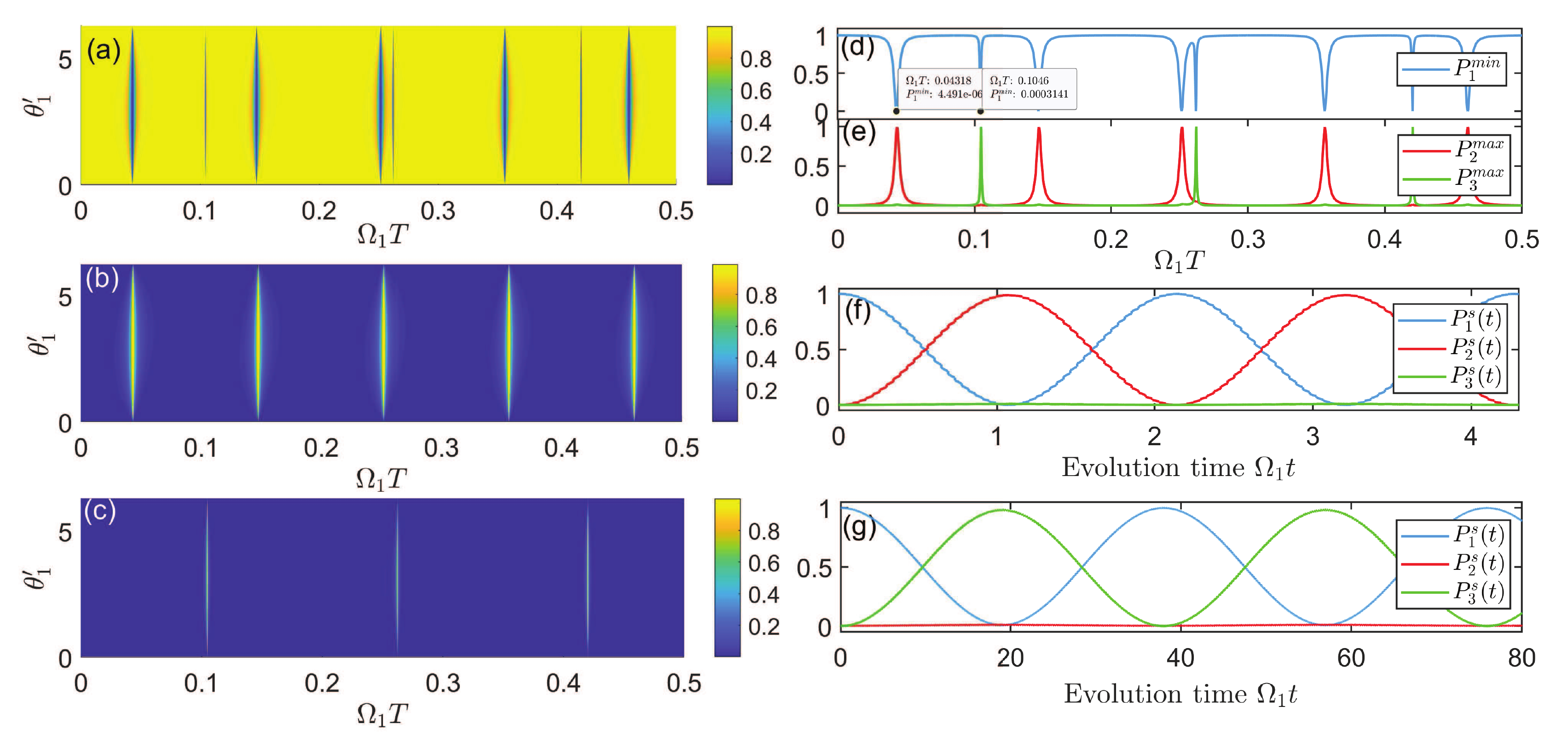}\\[5pt]  
\caption{  (color online) (a) $P_1^{\mathrm{min}}$, (b) $P_2^{\mathrm{max}}$, and (c) $P_3^{\mathrm{max}}$  vs $T$ and $\theta_1'$, where the parameters are given in Table~\ref{bt2}. (d) $P_1^{\mathrm{min}}$ and (e) $P_j^{\mathrm{max}}$ ($j=2,3$) vs $T$ by setting $\theta_1'=\pi$ in panels (a)-(c). (f) Population evolution of the state $|j\rangle$ $(j=1,2,3)$, where the system is driven into the one-photon resonance regime. (g) Population evolution of the state $|j\rangle$ $(j=1,2,3)$, where the system is driven into the two-photon resonance regime.   }\label{fig:S6}
\end{center}
\end{figure}

For the phase kick, we can modulate $\theta_1'$ or $\theta_2'$ while the other parameters remain unchanged, namely $\Omega_l'=\Omega_l$ and $\Delta_l'=\Delta_l$ ($l=1,2$). We illustrate this situation by periodically kicking the phase $\theta_1'$ in Fig.~\ref{fig:S6}. The relevant results are similar to the situation of the amplitude kick.
Furthermore, both Fig.~\ref{fig:S4} and Fig.~\ref{fig:S6} demonstrate that the dynamics of the three-level system are very versatile by altering the parameters of periodic kicks. In particular, one can modulate the coupling strength or the phase rather than the detuning (namely the frequency of control fields) to readily realize transformation between different regimes, such as from the one-photon resonance regime to the two-photon resonance regime.
Therefore, the coherent control technique would become more diverse when combined with periodic kicks.

\section{Applications of periodic kicks}  \label{iiia}

Obviously, the most direct application of periodic kicks is to employ the coherent destruction of coupling phenomenon as one of the control techniques to manipulate quantum states.
For example, in the system of the ultracold atoms trapped in a one-dimensional optical lattice,\ucite{PhysRevLett.99.110501} the sinusoidal driving field can be replaced by periodic kicks to implement quantum logic gates as well as the entanglement between distinct spatial locations.
The advantages are that there are several adjustable physical parameters in periodic kicks. In particular, the dynamical behaviors always preserve quantum coherence during the whole evolution process.

Another application of periodic kicks is to control system dynamics. To be specific, suppose that a quantum system to be controlled is initially in the large detuning regime. Thus, the evolution of system states is permanently frozen, even though this system interacts with control fields. According to the analysis in Sections.~2-3, when we impulse a certain number of additional kicks to the system, the dynamics would evolve. Then, we remove the kicks at some given moment, and the dynamics would be frozen again. By adding and removing the kicks, we achieve the goal of controlling system dynamics.

\begin{figure}
\begin{center}
\includegraphics[width=6in]{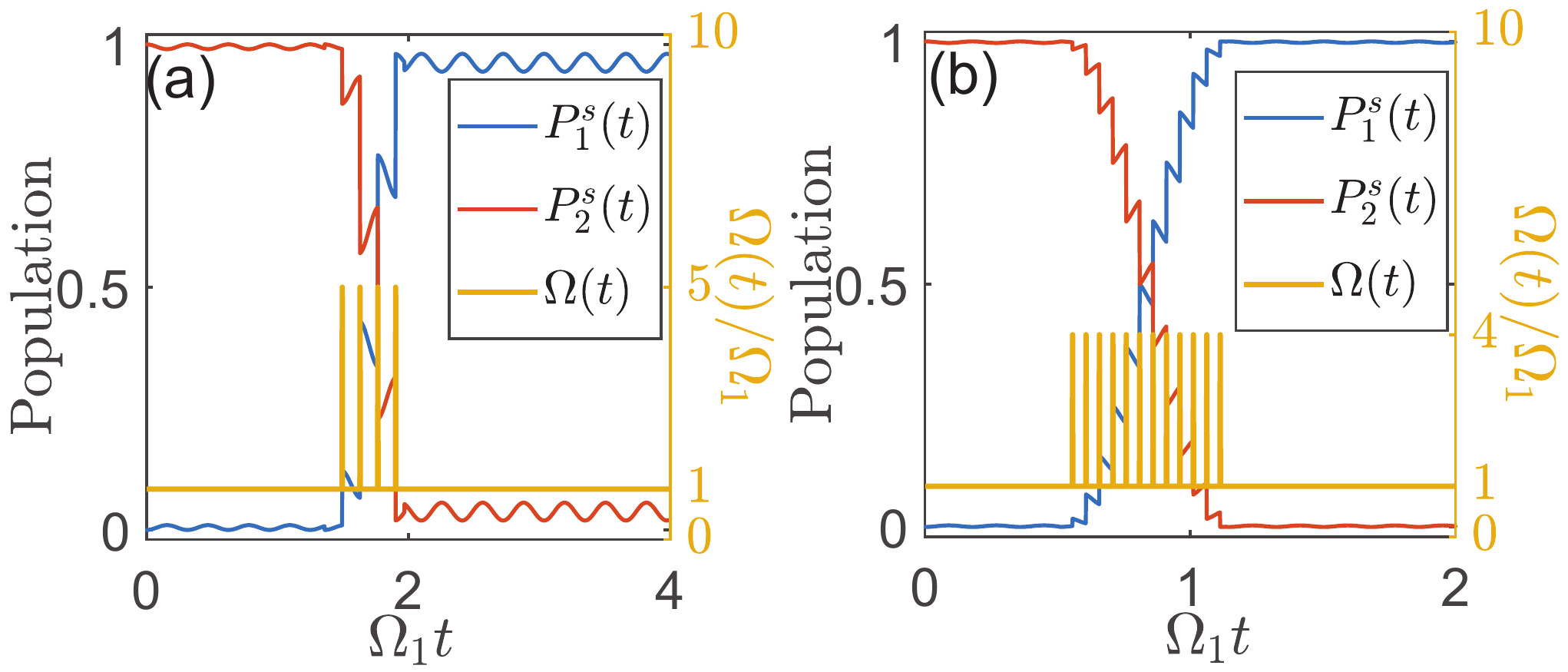}\\[5pt]  
\caption{  (color online) Time evolution of each state (the left-black vertical axis) and the coupling strength (the right-yellow vertical axis) in the two-level system by adding a certain number of kicks to the coupling strength. The parameters are given in Table~\ref{bt1}.
Population inversion is successfully obtained by only adding 12 kicks in panel (b). }\label{fig:03s}
\end{center}
\end{figure}

As the first example, we demonstrate how to achieve population inversion in a two-level system by periodic kicks, which is plotted in
Fig.~\ref{fig:03s}(a). We can observe that the population of the states $|1\rangle$ and $|2\rangle$ almost keep unchanged initially due to the large detuning regime. When we add kicks to the coupling strength, i.e.,
\begin{eqnarray}
\Omega(t)=\left \{
\begin{array}{ll}
    \Omega_1,  ~~~~~~t\neq nT, \\[1.1ex]
    \Omega_1', ~~~~~~t=nT, \\
\end{array}
\right.
\end{eqnarray}
the population of the states $|1\rangle$ and $|2\rangle$ begin to change. After 4 kicks, we almost achieve population inversion in this system. Then, we immediately remove the kicks, and the system dynamics recover to frozen again.
Note that there are still slight oscillations after removing the kicks.
The reason originates from two perspectives. (i) The detuning $\Delta_1$ of the original system is not large enough. (ii) The parameters of periodic kicks are chosen improperly so that the effective pulse number $N_{\mathrm{eff}}={\pi}/({2\Omega_{\mathrm{eff}}T})$ cannot be well regarded as an integer.
Once the system parameters are appropriately selected to satisfy the above perspectives, the population would be successfully inverse and nearly unchanged after removing the kicks. The relevant results are shown in Fig.~\ref{fig:03s}(b), where the oscillations virtually disappear after 12 kicks.

Another example is that one can implement selective transition in the same multilevel system by periodic kicks.
This operation is also achieved by adding and removing the kicks in the original system.
In the following, we illustrate this procedure in the three-level system.
First, we put the three-level system in the large detuning regime initially, i.e., $|\Delta_l|\gg\Omega_m$ ($l,m=1,2$). As a result, the population of the state $|j\rangle$ ($j=1,2,3$) cannot evolve. If we add the kicks to the coupling strength (phase) and the parameters of the kicks satisfy the relation shown in Fig.~\ref{fig:S4}(f) [Fig.~\ref{fig:S6}(f)], the system would be steered into the superposition states of $|1\rangle$ and $|2\rangle$.
Next, when obtaining the desired superposition state of $|1\rangle$ and $|2\rangle$, we directly remove the kicks, and the system states would remain unchanged since then.
On the other hand, if we want to prepare the superposition states of $|1\rangle$ and $|3\rangle$, we can add the kicks to the coupling strength (phase), and the parameters of the kicks should satisfy the relation shown in Fig.~\ref{fig:S4}(g) [Fig.~\ref{fig:S6}(g)].
It is worth mentioning that although we prepare different superposition states in the same three-level system, we do not alter the physical parameters except for the time interval (period) of the periodic kicks, and different periodic kicks would correspond to different transitions.

Note that one advantage of utilizing periodic kicks is flexibility and adjustability in a multilevel system.
For example, when the original three-level system is in the resonance regime, i.e., $\Delta_1=\Delta_2=0$, there are only two coupling strengths $\Omega_1$ and $\Omega_2$ to be modulated. However, when the original system is in the large detuning regime and we drive this system into the resonance regime by periodic kicks, all parameters (the period $T$, the coupling strengths $\Omega_j$, the detunings $\Delta_j$, and the phases $\theta_j$, $j=1,2$) can be used to modulate. Therefore, more adjustable parameters are involved in this system by periodic kicks.
Another advantage is that the population of the remaining states is sharply suppressed by periodic kicks. For example, when preparing the superposition states of $|1\rangle$ and $|2\rangle$ ($|3\rangle$), the population of $|3\rangle$ ($|2\rangle$) would not be excited during the whole process due to the large detuning regime, cf.~Fig.~\ref{fig:S4}(f) [Fig.~\ref{fig:S4}(g)].
Note that the interaction between the quantum system and control fields sometimes brings a slight change of resonance frequency of the system, which is known as the ``Stark shift'' in quantum optics.\ucite{scully97} Generally, one needs extra control fields to eliminate it.
However, it makes no difference in the system driven by periodic kicks, since the Stark shift can be blent in the detuning, and the detuning can be arbitrarily modulated in principle. This is also the advantage when controlling system dynamics by periodic kicks.

\section{Physical implementation of periodic kicks}  \label{5v}

Strictly speaking, periodic kicks do not exist in the physical system, since the instantaneous pulse with zero time duration is unattainable in experiments.
However, one can substitute periodic kicks for other easily accessible pulse shapes to obtain equivalent physical effects. In this section, we study this issue.

As shown in Eq.~(\ref{6}), the evolution operator governed by the kick Hamiltonian $H_1'(t)$ reads $U_1'=\exp(-i\mathcal{M}_1)$, where $\mathcal{M}_1=\Delta_1'|2\rangle\langle2| +\Omega_1'e^{i\theta_1'}|1\rangle\langle2|+\mathrm{H.c.}$
Suppose we have another Hamiltonian $\mathcal{H}_1'(t)$, and the evolution operator governed by this Hamiltonian is
\begin{eqnarray}
\mathcal{U}_1(t)=\mathbb{T}\exp\left[-i\int_{0}^{t}\mathcal{H}_1'(\tau)d\tau\right],
\end{eqnarray}
where $\mathbb{T}$ is the time-ordering operator. For the sake of simplicity, we consider the Hamiltonian $\mathcal{H}_1'(t)$ is time-independent, and thus $\mathcal{U}_1(t)=\exp\left(-i\mathcal{H}_1't\right)$. Then, at a particular evolution time $T'$, we demand that the following equation be satisfied
\begin{eqnarray}    \label{s18a}
\mathcal{U}_1(T')=\exp\left(-i\mathcal{H}_1'T'\right)=U_1'=\exp(-i\mathcal{M}_1),
\end{eqnarray}
which means that the kick dynamics is equivalent to the dynamics governed by the Hamiltonian $\mathcal{H}_1'$ at the evolution time $T'$ (up to a global phase).
As a result, we can safely replace the kick Hamiltonian $H_1'(t)$ with the Hamiltonian $\mathcal{H}_1'$.

Next, we demonstrate how to construct the Hamiltonian $\mathcal{H}_1'$.
One of the most intuitive ways can be found as follows. First, we diagonalize $\mathcal{M}_1$ to obtain the eigenstates $\{|E_n\rangle\}$, namely
\begin{eqnarray}
\mathcal{M}_1=\sum_{n=1}^{N}E_n'|E_n\rangle\langle E_n|.
\end{eqnarray}
Then, the Hamiltonian $\mathcal{H}_1'$ is constructed as
\begin{eqnarray}
\mathcal{H}_1'=\sum_{n=1}^{N}\mathcal{E}_n'|E_n\rangle\langle E_n|,
\end{eqnarray}
where the undetermined coefficients $\mathcal{E}_n'$ should satisfy
\begin{eqnarray}    \label{s20}
\mathcal{E}_n' T'+2k_{n}\pi=E_n',~~~n=1,2,\dots,N.
\end{eqnarray}
Here, $k_{n}$ can be chosen one of any integer, i.e., $k_{n}=0,\pm1,\pm2,\dots$.
Generally, the solution of the equations set~(\ref{s20}) always exists, since there are $2N+1$ variables (including the evolution time $T'$) while it has only $N$ equations.
Therefore, plenty of Hamiltonians can be employed to substitute for the kick Hamiltonian.

As an example, in the two-level system, we simply set $\mathcal{H}_1'=H_1'$, then the evolution time $T'$ is easily yielded by solving the equations set~(\ref{s20}), which reads
\begin{eqnarray}
T'=|1-\frac{2k\pi}{E_1'}|,~~~k=0,\pm1,\pm2,\dots
\end{eqnarray}
In this situation, periodic kicks become a periodic square-wave pulse, where the Hamiltonian $H_1$ lasts the duration $T$, the Hamiltonian $H_1'$ lasts the duration $T'$. Therefore, the period $T_s$ of the square-wave pulse reads $T_s=T+T'$, and the form of the system Hamiltonian can be written as
\begin{eqnarray}
H(t)=\left \{
\begin{array}{ll}
    H_1,  ~~~~~~t\in(mT_s, T+mT_s], \\[1.1ex]
    H_1', ~~~~~~t\in(T+mT_s, (m+1)T_s], \\
\end{array}
\right.
\end{eqnarray}
where $m\in\mathbb{N}$.
It is not hard to calculate the effective coupling strength $\Omega_{\mathrm{eff}}'$ under the periodic square-wave pulse, which is given by
\begin{eqnarray}
\Omega_{\mathrm{eff}}'=\frac{T}{T_s}\Omega_{\mathrm{eff}},
\end{eqnarray}
where $\Omega_{\mathrm{eff}}$ is the effective coupling strength under periodic kicks.

\begin{figure}
\begin{center}
\includegraphics[width=6in]{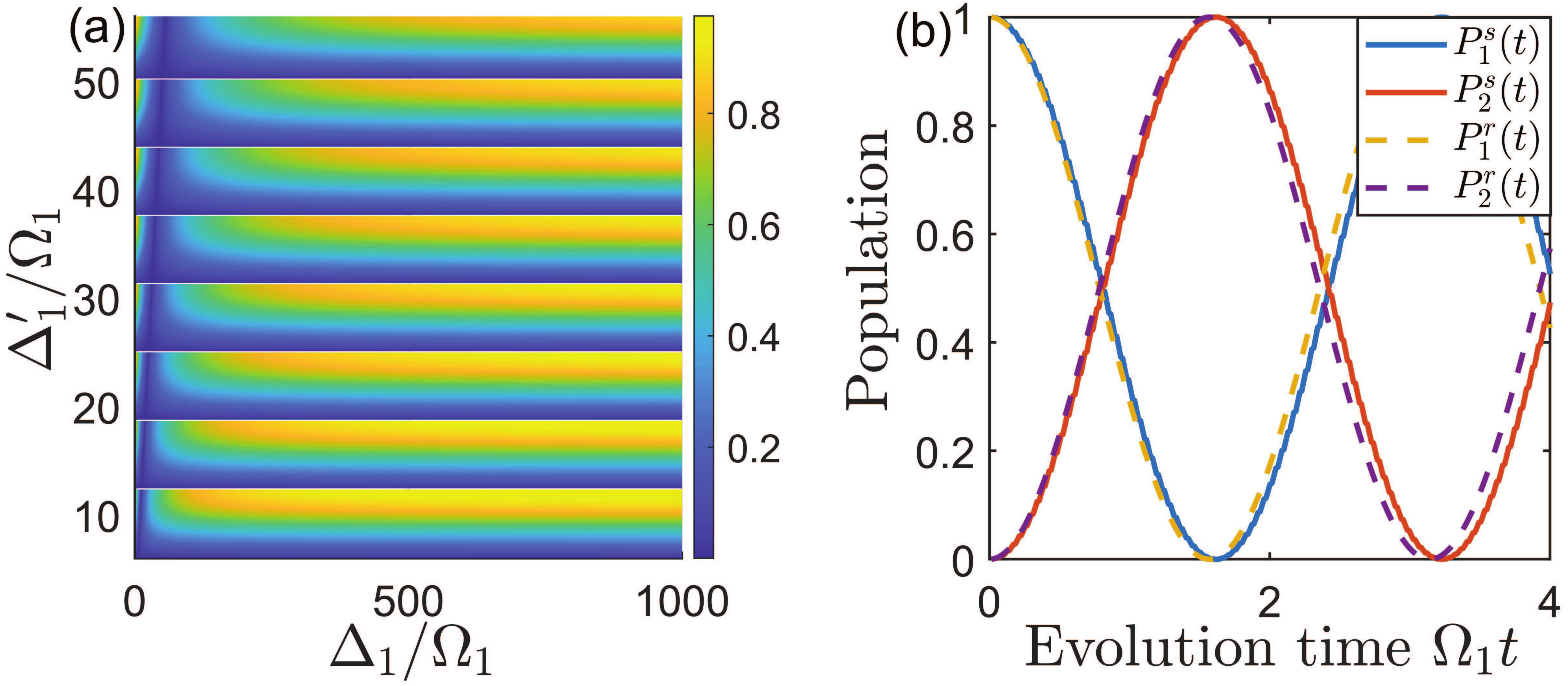}\\[5pt]  
\caption{  (color online) (a) $\Omega_{\mathrm{eff}}'$ vs $\Delta_1$ and $\Delta_1'$. In the periodic square-wave pulse, we only change the detuning and the other parameters remain unchanged, i.e., $\Omega_1'=\Omega_1$ and $\theta_1'=\theta_1=0$. (b)~Time evolution of states $|j\rangle$ ($j=1,2$) described by the periodic square-wave pulse (the solid curves) and the resonance pulse (the dash curves), where the parameters are $\Delta_1/\Omega_1=1000$ and $\Delta_1'/\Omega_1=12.3$.} \label{fig:S3}
\end{center}
\end{figure}

In Fig.~\ref{fig:S3}(a), we plot the effective coupling strength $\Omega_{\mathrm{eff}}'$ as a function of the detunings $\Delta_1$ and $\Delta_1'$ under the periodic square-wave pulse. The results demonstrate that unlike the periodic kicks case, the effective coupling strength $\Omega_{\mathrm{eff}}'$ cannot exceed the coupling strength $\Omega_1$ of the original system, and it always exists the following inequation
\begin{eqnarray}
\Omega_{\mathrm{eff}}'<\Omega_1.
\end{eqnarray}
It is further shown in Fig.~\ref{fig:S3}(a) that $\Omega_{\mathrm{eff}}'$ can be nearly close to $\Omega_1$ if $\Delta_1$ is large enough.
Therefore, we can design different periodic square-wave pulses whose dynamics are almost identical to the resonance pulse situation, a special case ($\Omega_{\mathrm{eff}}'\approx0.9753\Omega_1$) shown in Fig.~\ref{fig:S3}(b).
As a comparison, Fig.~\ref{fig:S3}(b) demonstrates the population evolution of the state $|j\rangle$ ($j=1,2$) described by the periodic square-wave pulse (the solid curves) and the resonance pulse (the dash curves) respectively, where the Hamiltonian of the resonance pulse reads
\begin{eqnarray}   \label{24}
H_r=\Omega_1|1\rangle\langle2|+\Omega_1|2\rangle\langle1|.
\end{eqnarray}

Note that although the resonance pulse is the fastest dynamical evolution, there is only a single adjustable parameter $\Omega_1$.
As a result, the system dynamics are sensitive to external noises induced by $\Omega_1$.
Nevertheless, we find that the dynamics of the resonance pulse can almost be replaced by the periodic square-wave pulse, as shown in Fig.~\ref{fig:S3}(b).
The advantage of this substitution is that we can control the system dynamics by modulating different physical parameters now. At the same time, one can design the waveform to be immune to some type of noise. Above all, the evolution time is almost close to the shortest.

\section{Conclusion}  \label{v}

We have studied the kick dynamics of quantum systems and given the analytical expression of the time-independent effective Hamiltonian, which describes the dynamical behavior very well in a long timescale.
Notably, the effective Hamiltonian becomes more valid when the two-level system is in the large detuning regime.
By choosing suitable parameters of periodic kicks, we would observe the coherent destruction of coupling phenomenon in this system, which makes the evolution of quantum states freeze again even in the resonance regime ($\Delta_{\mathrm{eff}}=0$).
In addition, the effective coupling strength can be much larger than the coupling strength of the original system in the frequency kick and phase kick, which stems from the instantaneous of kicks, namely zero time duration.

Different from the two-level system, we have further shown that the dynamics exhibit more variation in the three-level system.
The same three-level system can stay in different regimes under different parameters of periodic kicks.
More importantly, by merely modulating a single parameter of periodic kicks, we readily switch between those regimes, such as from the original large detuning regime to the resonance regime ($\Delta_{\mathrm{eff},1}=\Delta_{\mathrm{eff},2}=0$), the one-photon resonance regime ($\Delta_{\mathrm{eff},1}=0$), or the two-photon resonance regime ($\Delta_{\mathrm{eff},2}=0$).

Due to the prominent properties of periodic kicks, we can easily control quantum states as desired.
For its application, we have concerned with three aspects: (i) Manipulate quantum states with the help of the coherent destruction of coupling phenomenon; (ii) Control system dynamics by adding a certain number of kicks, e.g., achieving population inversion or qubit operations; (iii) Implement quantum state selective transition while sharply suppressing the population of remainder excited states in three-level systems. Finally, we have provided the experimental feasible pulses as a substitution for periodic kicks.
In a word, periodic kicks can be applied in a wide variety of quantum systems due to their flexibility and adjustability.
Thus, it is believed that periodic kicks would offer us a novel method of coherent control in quantum information processing.

\section*{Appendix A: Zeros of the function $f_2$}   \label{IIa}

As claimed in the main text, whether the system is in the resonance regime ($\Delta_{\textrm{eff}}=0$) is only determined by the function $f_2$.
In this Appendix, we study the roots of the function $f_2$ in detail.
Since there are four variables $\{T,\Omega_1',\Delta_1',\theta_1'\}$ in the function $f_2$,
we investigate these situations individually in the following. We adopt a convention that the variable appears in the bracket of the function $f_2$ to distinguish different situations.

(i) Provided that $T$ is variable and $\{\Omega_1',\theta_1',\Delta_1'\}$ are arbitrary constants. It is not hard to calculate that
\begin{eqnarray}
f_2(T)\!\!\!&=&\!\!\!\frac{\Delta_1}{2E_1}\cos E_1'\sin E_1T+\frac{\Delta_1'}{2E_1'}\sin E_1'\cos E_1T+\frac{\Omega_1\Omega_1'\sin(\theta_1-\theta_1')}{E_1E_1'}\sin E_1' \sin E_1T  \nonumber\\[0.5ex]
\!\!\!&=&\!\!\!A_1\sin(E_1T+\phi_1),
\end{eqnarray}
where
\begin{eqnarray}
A_1\!\!\!&=&\!\!\!\sqrt{\left[\frac{\Delta_1}{2E_1}\cos E_1'+\frac{\Omega_1\Omega_1'\sin(\theta_1-\theta_1')}{E_1E_1'}\sin E_1'\right]^2+\left(\frac{\Delta_1'}{2E_1'}\sin E_1'\right)^2}, \\[0.5ex]
\phi_1\!\!\!&=&\!\!\!\arctan\frac{\Delta_1'E_1\sin E_1'}{\Delta_1E_1'\cos E_1'+2\Omega_1\Omega_1'\sin(\theta_1-\theta_1')\sin E_1'}.
\end{eqnarray}
We readily find that the roots of the equation $f_2(T)=0$ always exist, which read
\begin{eqnarray}
T=\frac{-\phi_1+n\pi}{E_1},~~~ n=1,2,3, \dots.
\end{eqnarray}

(ii) Provided that $\theta_1'$ is variable and $\{T, \Omega_1',\Delta_1'\}$ are arbitrary constants. It is not hard to calculate that
\begin{eqnarray}
f_2(\theta_1')\!\!\!&=&\!\!\!\frac{\Delta_1}{2E_1}\cos E_1'\sin E_1T+\frac{\Delta_1'}{2E_1'}\sin E_1'\cos E_1T+\frac{\Omega_1\Omega_1'\sin(\theta_1-\theta_1')}{E_1E_1'}\sin E_1' \sin E_1T    \nonumber\\[0.5ex]
\!\!\!&=&\!\!\!A_2\sin(E_1T+\phi_2)+\frac{\Omega_1\Omega_1'\sin(\theta_1-\theta_1')}{E_1E_1'}\sin E_1' \sin E_1T,
\end{eqnarray}
where
\begin{eqnarray}
A_2=\sqrt{\left(\frac{\Delta_1}{2E_1}\cos E_1'\right)^2+\left(\frac{\Delta_1'}{2E_1'}\sin E_1'\right)^2},~~~~~~~~~ \phi_2=\arctan\frac{\Delta_1'E_1\tan E_1'}{\Delta_1E_1'}.
\end{eqnarray}
We can find that the maximum of the first term is $A_2\geq \min\{\frac{\Delta_1}{2E_1}, \frac{\Delta_1'}{2E_1'}\}$,
and the minimum of the second term is $\frac{\Omega_1\Omega_1'}{E_1E_1'}\leq\max\{\frac{\Omega_1}{E_1}, \frac{\Omega_1'}{E_1'}\}$.
In the large detuning regime ($\Delta_k\gg\Omega_k, k=1,2)$, the inequation $A_2\gg\frac{\Omega_1\Omega_1'}{E_1E_1'}$ is always satisfied. Hence, $f_2(\theta_1')\gg0$ for arbitrary $\theta_1'$.
As a result, the zeros of the function $f_2(\theta_1')$ do not always exist when $\{T, \Omega_1',\Delta_1'\}$ are arbitrary constants.
If and only if $\{T, \Omega_1',\Delta_1'\}$ satisfy the inequation:
\begin{eqnarray}
\frac{E_1E_1'A_2\sin(E_1T+\phi_2)}{\Omega_1\Omega_1'\sin E_1' \sin E_1T}\leq1,
\end{eqnarray}
the zeros of the function $f_2(\theta_1')$ exist, and the form reads
\begin{eqnarray}
\theta_1'=\theta_1+\arcsin\frac{E_1E_1'A_2\sin(E_1T+\phi_2)}{\Omega_1\Omega_1'\sin E_1' \sin E_1T}+2n\pi, ~~~~~n=1,2,3, \dots.
\end{eqnarray}

\begin{figure}
\begin{center}
\includegraphics[width=6in]{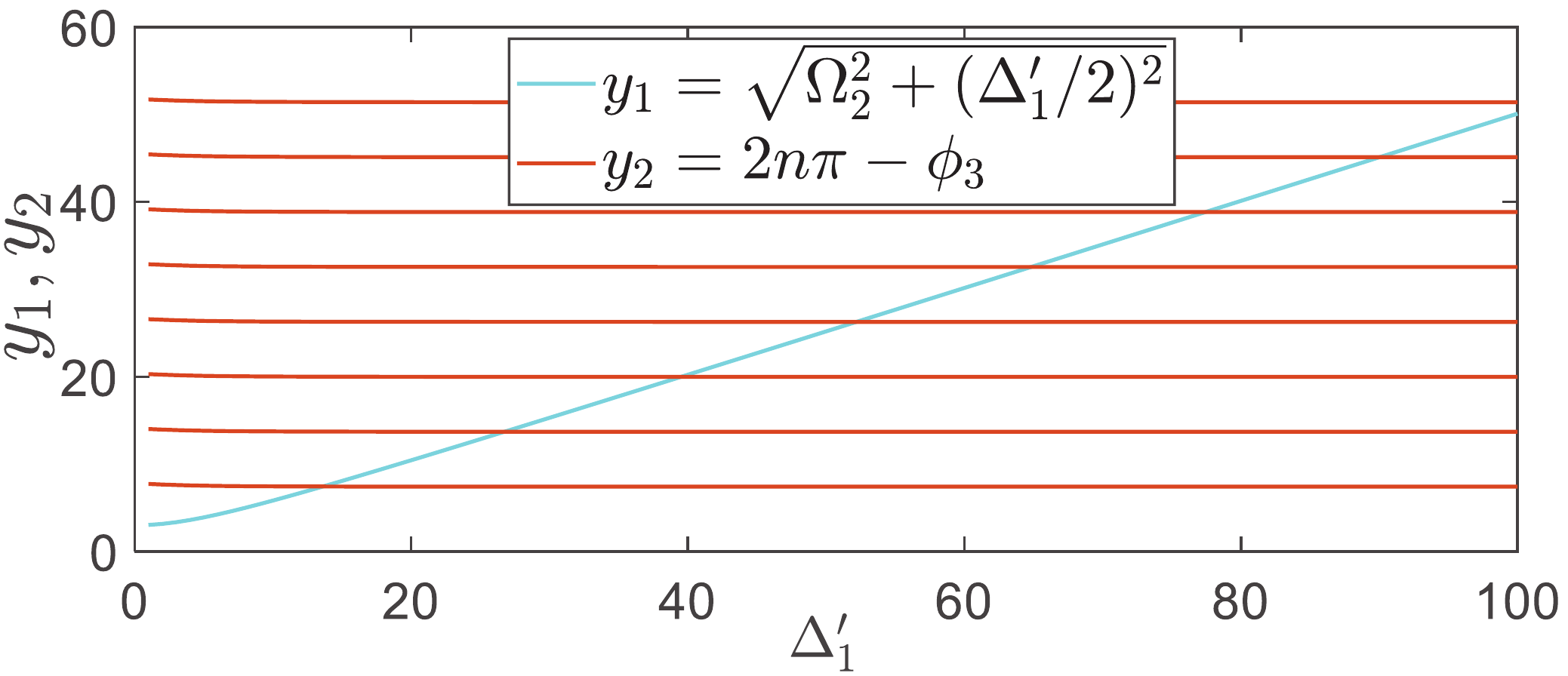}\\[5pt]  
\caption{  (color online) The curves $y_1$ and $y_2$ as a function of $\Delta_1'$, where $\Delta_1=40$, $\Omega_1=1$, $\theta_1=0$, $\Omega_1'=3$, and $\theta_1'=\pi/3$.}\label{fig:S1}
\end{center}
\end{figure}

(iii) Provided that $\Delta_1'$ is variable and $\{T, \Omega_1',\theta_1'\}$ are arbitrary constants. It is not hard to calculate that
\begin{eqnarray}
f_2(\Delta_1')\!\!\!&=&\!\!\!\frac{\Delta_1}{2E_1}\cos E_1'\sin E_1T+\frac{\Delta_1'}{2E_1'}\sin E_1'\cos E_1T+\frac{\Omega_1\Omega_1'\sin(\theta_1-\theta_1')}{E_1E_1'}\sin E_1' \sin E_1T \nonumber\\[0.5ex]
\!\!\!&=&\!\!\!A_3\sin(E_1'+\phi_3),
\end{eqnarray}
where
\begin{eqnarray}
A_3\!\!\!&=&\!\!\!\sqrt{\left[\frac{\Delta_1'}{2E_1'}\cos E_1T+\frac{\Omega_1\Omega_1'\sin(\theta_1-\theta_1')}{E_1E_1'} \sin E_1T\right]^2+\left(\frac{\Delta_1}{2E_1}\sin E_1T\right)^2},  \\[0.5ex]
\phi_3\!\!\!&=&\!\!\!\arctan\frac{\Delta_1E_1'\sin E_1T}{\Delta_1'E_1\cos E_1T+2\Omega_1\Omega_1'\sin(\theta_1-\theta_1')\sin E_1T}.
\end{eqnarray}
In order to satisfy the equation $f_2(\Delta_1')=0$, we only require to satisfy the equation $E_1'+\phi_3=n\pi$. Generally speaking, the transcendental equation $E_1'+\phi_3=n\pi$ does not have the analytical solution. Here, we can employ graphical method to confirm whether this transcendental equation is valid or not.
The first curve $y_1(\Delta_1')=E_1'=\sqrt{\Omega_1'^2+(\Delta_1'/2)^2}$ is a hyperbolic curve, whose focus is on the $y$-axis, while the second curve $y_2(\Delta_1')=n\pi-\phi_3$ contain $n$ inverse tangent curves.
There must be points of intersection between the curves $y_1$ and $y_2$, as shown in Fig.~\ref{fig:S1}. Hence, the zeros of the function $f_2(\Delta_1')$ always exist.

(iv) Provided that $\Omega_1'$ is variable and $\{T, \Delta_1',\theta_1'\}$ are arbitrary constants.
This situation is similar to the situation (iii), and we do not demonstrate here again. As a result, the zeros of the function $f_2(\Omega_1')$ always exist.

Therefore, not all parameters $\{T,\Omega_1',\Delta_1',\theta_1'\}$ can be individually modulated to guarantee the equation $f_2=0$, namely $\Delta_{\textrm{eff}}=0$.

\section*{Appendix B: Limitation of the effective coupling strength $\Omega_{\textrm{eff}}$ at $E_1'\rightarrow(m\pi)^{-}$}  \label{IIIa}

In this appendix, we first calculate the effective coupling strength when $E_1'=m\pi, m\in\mathbb{N}$. Since $E_1'=m\pi$, we can find that $\sin E_1'=0$ and $\cos E_1'=1$. Then, according to the equation $f_2=0$, it is easily found that $\sin E_1T=0$ and $\cos E_1T=1$. Substituting them into the other three functions, we obtain that $f_1=1$, $f_3=0$, and $f_4=0$. As a result, the evolution operator of this system within one period reads
\begin{eqnarray}
U(T)&=&\left(
                \begin{array}{cc}
                  1 & 0 \\
                  0 & 1 \\
                \end{array}
              \right).
\end{eqnarray}
Then, by inversely solving the equation $U(T)=e^{-iH_{\textrm{eff}}T}$, we find that the effective coupling strength $\Omega_{\textrm{eff}}=0$.

Next, we calculate the value of the effective coupling strength $\Omega_{\textrm{eff}}$ at limitation $E_1'\rightarrow(m\pi)^{-}$.
Note that the following equations are always satisfied:
\begin{eqnarray}
&&\lim\limits_{{E_1'\rightarrow(2m\pi)^{-}}}\sin E_1'=\lim\limits_{{x\rightarrow0^{-}}}\sin x,  \\\cr
&&\lim\limits_{{E_1'\rightarrow(2m\pi)^{-}}}\cos E_1'=\lim\limits_{{x\rightarrow0^{-}}}\cos x,  \\\cr
&&\lim\limits_{{E_1'\rightarrow(2m\pi)^{-}}}\arctan\frac{\Delta_1'E_1\sin E_1'}{\Delta_1E_1'\cos E_1'+2\Omega_1\Omega_1'\sin(\theta_1-\theta_1')} =\lim\limits_{{x\rightarrow0^{-}}}\arctan\frac{\Delta_1'E_1\sin x}{\Delta_1E_1'\cos x+2\Omega_1\Omega_1'\sin(\theta_1-\theta_1')}<0.~~~~~~
\end{eqnarray}
From the expressions $\Omega_{\textrm{eff}}=({\arccos|f_1|})/{T}$ and $T=({-\phi_1+n\pi})/{E_1}$, we find that the period $T={-\phi_1}/{E_1}$ if $\phi_1<0$ while the period $T=({-\phi_1+\pi})/{E_1}$ if $\phi_1>0$ in order to achieve the maximum value of $\Omega_{\textrm{eff}}$.
As a result, $E_1T=-\phi_1$ in limit of $E_1'\rightarrow(2m\pi)^{-}$.
Then, we rewrite the expression of the effective coupling strength as
\begin{eqnarray}  \label{b3}
\lim_{{E_1'\rightarrow(2m\pi)^{-}}}\Omega_{\textrm{eff}}&=& \lim_{{E_1'\rightarrow(2m\pi)^{-}}}\frac{\arccos|f_1|}{T}=\lim_{{x\rightarrow 0^{-}}}-E_1\frac{\sqrt{1-f_1^2}}{\phi_1}=\lim_{{x\rightarrow 0^{-}}}E_1\sqrt{\frac{2(1-f_1)}{\phi_1^2}}  \nonumber\\[1ex]
&=&\lim_{{x\rightarrow 0^{-}}}\frac{\Delta_1E_1'+2\Omega_1\Omega_1'\sin(\theta_1-\theta_1')}{\Delta_1'}\sqrt{\frac{2-2\left[\cos x\cos \phi_1+\frac{\Delta_1\Delta_1'+4\Omega_1\Omega_1'\cos(\theta_1-\theta_1')}{4E_1E_1'}\sin x \sin \phi_1\right]}{x^2}}  \nonumber\\[1ex]
&=&\lim_{{x\rightarrow 0^{-}}}\frac{\Delta_1E_1'+2\Omega_1\Omega_1'\sin(\theta_1-\theta_1')}{\Delta_1'} \sqrt{\frac{x^2+\phi_1^2-\frac{\Delta_1\Delta_1'+4\Omega_1\Omega_1'\cos(\theta_1-\theta_1')} {4E_1E_1'}x\phi_1}{x^2}}  \nonumber\\[1ex]
&=&\sqrt{|\Omega_1-\frac{\Delta_1}{\Delta_1'}\Omega_1'e^{i(\theta_1-\theta_1')}|^2 +f_5 \sin(\theta_1-\theta_1')},
\end{eqnarray}
where
\begin{eqnarray}
f_5=\frac{4\Omega_1\Omega_1'^2[\Omega_1E_1'\sin(\theta_1-\theta_1')- \Omega_1\Delta_1'\cos(\theta_1-\theta_1')+\Delta_1\Omega_1']}{E_1'\Delta_1'^2},
\end{eqnarray}
and we have used the series expanded of $\phi_1$:
\begin{eqnarray}
\phi_1=\frac{\Delta_1'E_1}{\Delta_1E_1'+2\Omega_1\Omega_1'\sin(\theta_1-\theta_1')}x+\mathcal{O}(x^2).
\end{eqnarray}

Therefore, according to Eq.~(\ref{b3}), in the frequency kick, the effective coupling strength becomes
\begin{eqnarray}
\lim\limits_{{E_1'\rightarrow(2m\pi)^{-}}}\Omega_{\textrm{eff}}=\Omega_1|1-\frac{\Delta_1}{\Delta_1'}|.
\end{eqnarray}
In the amplitude kick, the effective coupling strength becomes
\begin{eqnarray}
\lim\limits_{{E_1'\rightarrow(2m\pi)^{-}}}\Omega_{\textrm{eff}}=|\Omega_1-\Omega_1'|.
\end{eqnarray}
In the phase kick, the effective coupling strength can be estimated as in the large detuning regime:
\begin{eqnarray}
\lim\limits_{{E_1'\rightarrow(2m\pi)^{-}}}\Omega_{\textrm{eff}} \approx|\Omega_1-\frac{\Delta_1}{\Delta_1'}\Omega_1'e^{i(\theta_1-\theta_1')}|.
\end{eqnarray}
Besides, we can calculate the value of the effective coupling strength when $E_1'\rightarrow[(2m+1)\pi]^{-}$ and the similar results would be obtained.

\section*{Appendix C: Deviation of the effective Hamiltonian of the three-level system driven by periodic kicks}  \label{IVa}

In this appendix, we mainly study the kick dynamics in the three-level system. The Hamiltonian of the original three-level system reads
\begin{eqnarray}   \label{s19}
H_1=\Delta_1|2\rangle\langle 2|+2\Delta_1|3\rangle\langle 3|+\Omega_1e^{i\theta_1}|1\rangle\langle 2|+\Omega_1e^{i\theta_1}|2\rangle\langle 3|+\mathrm{H.c.},
\end{eqnarray}
where $\Delta_1$, $\Omega_1$, and $\theta_1$ represent the detuning, the coupling strength, and the phase, respectively.
It is not hard to calculate the evolution operator of three-level system and the matrix form in basis $\{|1\rangle, |2\rangle, |3\rangle\}$ reads
\begin{eqnarray}  \label{47s}
U_1(t)\!=\!\frac{e^{-i\Delta_1t}}{E_1^2}\!\left(
                    \begin{array}{ccc}
                      u_{11} & u_{12} & u_{13} \\
                      u_{21} & u_{22} & u_{23} \\
                      u_{31} & u_{32} & u_{33} \\
                    \end{array}
                  \right),
\end{eqnarray}
where $E_1=\sqrt{\Delta_1^2+2\Omega_1^2}$, and
\begin{eqnarray}
u_{11}\!\!\!&=&\!\!\!u^*_{33}=\Omega_1^2+(\Delta_1^2+\Omega_1^2)\cos E_1t+i\Delta_1E_1\sin E_1t,  \nonumber\\[0.5ex]
u_{22}\!\!\!&=&\!\!\!\Delta_1^2+2\Omega_1^2\cos E_1t,   \nonumber\\[0.5ex]
u_{12}\!\!\!&=&\!\!\!\Omega_1e^{i\theta_1}(\Delta_1-\Delta_1\cos E_1t-iE_1\sin E_1t),  \nonumber\\[0.5ex]
u_{21}\!\!\!&=&\!\!\!\Omega_1e^{-i\theta_1}(\Delta_1-\Delta_1\cos E_1t-iE_1\sin E_1t),  \nonumber\\[0.5ex]
u_{13}\!\!\!&=&\!\!\!u^*_{31}=\Omega_1^2e^{i2\theta_1}(\cos E_1t-1)  \nonumber\\[0.5ex]
u_{23}\!\!\!&=&\!\!\!\Omega_1e^{i\theta_1}(\Delta_1\cos E_1t-\Delta_1-iE_1\sin E_1t),  \nonumber\\[0.5ex]
u_{32}\!\!\!&=&\!\!\!\Omega_1e^{-i\theta_1}(\Delta_1\cos E_1t-\Delta_1-iE_1\sin E_1t).  \nonumber
\end{eqnarray}
Obviously, the population of the system states $\{|1\rangle, |2\rangle, |3\rangle\}$ are freezing and the dynamics cannot evolve in the large detuning regime ($|\Delta_1|\gg\Omega_1$).

Now, we turn to study the system dynamics under periodic kicks, and the Hamiltonian reads
\begin{eqnarray}
H(t)=H_1+H_1'(t),
\end{eqnarray}
where the kick Hamiltonian $H_1'(t)$ reads
\begin{eqnarray}
H_1'(t)=\sum_{n=-\infty}^{+\infty}\delta(t-nT)\left[\Delta_1'|2\rangle\langle 2|+2\Delta_1'|3\rangle\langle 3|+\Omega_1'e^{i\theta_1'}(|1\rangle\langle 2|+|2\rangle\langle 3|)+\mathrm{H.c.}\right].
\end{eqnarray}
Similarly to the case of the two-level system, the evolution operator of this system within one period can be written as
\begin{eqnarray}  \label{c5}
U(T)=e^{-i\mathcal{M}_1}e^{-iH_1T}=\frac{e^{-i\Delta_1'}e^{-i\Delta_1T}}{E_1'^2E_1^2}\left(
                    \begin{array}{ccc}
                       g_1(T)+ig_2(T) & g_8(T)+ig_9(T) &  g_5(T)-ig_6(T) \\
                       g_3(T)+ig_4(T) & g_7(T) & -g_3(T)+ig_4(T) \\
                       g_5(T)+ig_6(T) & -g_8(T)+ig_9(T) & g_1(T)-ig_2(T) \\
                    \end{array}
                  \right),
\end{eqnarray}
where $\mathcal{M}_1=\Delta_1'|2\rangle\langle 2|+2\Delta_1'|3\rangle\langle 3|+\Omega_1'e^{i\theta_1'}(|1\rangle\langle 2|+|2\rangle\langle 3|)+\mathrm{H.c.}$, $E_1'=\sqrt{\Delta_1'^2+2\Omega_1'^2}$, and the form of the function $g_k(T)$ reads
\begin{eqnarray}
g_k(T)=a_{k1}\cos E_1T+a_{k2}\sin E_1T+a_{k3}, ~~~~~k=1,\dots,9.
\end{eqnarray}
{Since the expressions of the coefficients $a_{kj}$ ($j=1,2,3$) can be easily derived from Eq.~(\ref{47s}), we do not intend to present their expressions here due to complexity.}

In principle, with the help of the expression of the evolution operator within one period, as shown in Eq.~(\ref{c5}), we can acquire the kick dynamics of the three-level system. However, what we are interested in here is to modulate the period $T$ of periodic kicks to drive the three-level system into the resonance regime (or equivalently, the effective detunings $\Delta_{\mathrm{eff},1}=\Delta_{\mathrm{eff},2}=0$).
That is, the kick dynamics of the three-level system is described by the following effective Hamiltonian
\begin{eqnarray}
H_{\textrm{eff}}=\Omega_{\textrm{eff}}e^{i\theta_{\textrm{eff}}}(|1\rangle\langle 2|+|2\rangle\langle 3|)+\mathrm{H.c.},
\end{eqnarray}
where the unknown coefficients $\Omega_{\textrm{eff}}$ and $\theta_{\textrm{eff}}$ represent the effective coupling strength and phase, respectively.
After some algebraic operations, the expression of period $T$ reads
\begin{eqnarray}
T=\frac{1}{E_1}\left({\arcsin\frac{-a_{23}}{\sqrt{a_{21}^2+a_{22}^2}}-\arctan\frac{a_{22}}{a_{21}}+2n\pi}\right),
\end{eqnarray}
or
\begin{eqnarray}
T=\frac{1}{E_1}\left[{\arcsin\frac{-a_{23}}{\sqrt{a_{21}^2+a_{22}^2}}-\arctan\frac{a_{22}}{a_{21}}+(2n+1)\pi}\right],
\end{eqnarray}
where $n=0,1,2,\dots$.
Then, by inversely solving the equation $U(T)=e^{-iH_{\textrm{eff}}T}$, we can obtain the effective coupling strength and phase:
\begin{eqnarray}
\Omega_{\textrm{eff}}=\frac{\sqrt{2}}{T}\arccos\frac{\sqrt{g_1(T)}}{E_1E_1'}, ~~~~~ \theta_{\textrm{eff}}=\arctan\frac{g_3(T)}{g_4(T)}.
\end{eqnarray}
Note that in order to achieve the above effective Hamiltonian, the period $T$ should examine whether it satisfies the following equations:
\begin{eqnarray}
&&g_3(T)+g_8(T)=g_4(T)-g_9(T)=2g_3(T)g_4(T)g_6(T)-g_5(T)[g_4^2(T)-g_3^2(T)]=0,  \\[0.5ex]
&&g_1(T)+\sqrt{g_5^2(T)+g_6^2(T)}=2g_1(T)-g_7(T)=1.
\end{eqnarray}

\section*{Acknowledgment}
We thank Shi-Biao Zheng for helpful discussions. This work is supported by National Natural Science Foundation of China (Grant Nos.~11805036, 12175033, 12147206), the Natural Science Foundation of Fujian Province (Grant No.~2021J01575), the Natural Science Funds for Distinguished Young Scholar of Fujian Province (Grant No.~2020J06011), and the Project from Fuzhou University (Grant No.~JG202001-2).

\end{document}